\documentclass[fleqn,usenatbib]{mnras}

\usepackage{newtxtext,newtxmath}

\usepackage[T1]{fontenc}
\usepackage{footmisc}
\DeclareRobustCommand{\VAN}[3]{#2}
\let\VANthebibliography\thebibliography
\def\thebibliography{\DeclareRobustCommand{\VAN}[3]{##3}\VANthebibliography}

\usepackage{graphicx}	
\usepackage{amsmath}	

\usepackage{float}
\usepackage{appendix}
\usepackage{xcolor}
\usepackage{amsmath}
\usepackage[normalem]{ulem}
\usepackage{changes}

\defcitealias{2023ApJ...954...63R}{Paper~I}

\title[97-min Eclipsing Period Bouncer]{A Joint SRG/eROSITA + ZTF Search: Discovery of a 97-min Period Eclipsing Cataclysmic Variable with Evidence of a Brown Dwarf Secondary}

\author[Galiullin and Rodriguez et al.]{
Ilkham Galiullin,$^{1}$\thanks{E-mail: IlhIGaliullin@kpfu.ru}
Antonio C. Rodriguez,$^{2}$
Shrinivas R. Kulkarni,$^{2}$
Rashid Sunyaev,$^{3,4,5}$
Marat Gilfanov,$^{3,4}$
\newauthor
Ilfan Bikmaev,$^{1,6}$
Lev Yungelson,$^{7}$
Jan van Roestel,$^{8}$
Boris T. G\"ansicke,$^{9}$
Irek Khamitov,$^{1,6}$
Paula Szkody,$^{10}$
\newauthor
Kareem El-Badry,$^{2}$
Mikhail Suslikov,$^{1}$
Thomas A. Prince,$^{11}$
Mikhail Buntov,$^{3}$
Ilaria Caiazzo,$^{2}$
Mark Gorbachev,$^{1}$
\newauthor
Matthew J. Graham,$^{2}$
Rustam Gumerov,$^{1,6}$
Eldar Irtuganov,$^{1}$
Russ R. Laher,$^{12}$
Pavel Medvedev,$^{3}$
Reed Riddle,$^{2}$
\newauthor
Ben Rusholme,$^{12}$
Nail Sakhibullin,$^{1,6}$
Alexander Sklyanov,$^{1}$ 
Zachary P. Vanderbosch$^{2}$
\\
$^{1}$ Kazan Federal University, Kremlevskaya Str.18, 420008, Kazan, Russia\\
$^{2}$ Department of Astronomy, California Institute of Technology, 1200 E. California Blvd, Pasadena, CA, 91125, USA\\
$^{3}$ Space Research Institute, Russian Academy of Sciences, Profsoyuznaya 84/32, 117997 Moscow, Russia\\
$^{4}$ Max Planck Institute for Astrophysics, Karl-Schwarzschild-Str 1, Garching b. Muenchen D-85741, Germany \\
$^{5}$ Institute for Advanced Study, 1st Einstein Drive, Princeton, NJ, 08540, USA\\
$^{6}$ Academy of Sciences of Tatarstan Rep., Baumana Str. 20, Kazan 420111, Russia \\
$^{7}$ Institute of Astronomy, Russian Academy of Sciences, 48 Pyatnitskaya str., Moscow 109017, Russia \\
$^{8}$ Anton Pannekoek Institute for Astronomy, University of Amsterdam, 1090 GE Amsterdam, The Netherlands \\
$^{9}$ Department of Physics, University of Warwick, Coventry CV4 7AL, UK \\
$^{10}$ Department of Astronomy, University of Washington, 3910 15th Avenue NE, Seattle, WA 98195, USA  \\
$^{11}$ Division of Physics, Mathematics, and Astronomy, California Institute of Technology, Pasadena, CA 91125, USA \\
$^{12}$ IPAC, California Institute of Technology, 1200 E. California Blvd, Pasadena, CA 91125, USA \\
}

\date{Accepted XXX. Received YYY; in original form ZZZ}

\pubyear{2023}

\begin{document}
\label{firstpage}
\pagerange{\pageref{firstpage}--\pageref{lastpage}}
\maketitle

\begin{abstract}
Cataclysmic variables (CVs) that have evolved past the period minimum during their lifetimes are predicted to be systems with a brown dwarf donor. While population synthesis models predict that around $\approx$40--70\% of the Galactic CVs are post-period minimum systems referred to as "period bouncers", only a few dozen confirmed systems are known. We report the study and characterisation of a new eclipsing CV, SRGeJ041130.3+685350 (SRGeJ0411), discovered from a joint SRG/eROSITA and ZTF program. The optical spectrum of SRGeJ0411 shows prominent hydrogen and helium emission lines, typical for CVs. We obtained optical high-speed photometry to confirm the eclipse of SRGeJ0411 and determine the orbital period to be $P_\textrm{orb} \approx 97.530$ minutes. The spectral energy distribution suggests that the donor has an effective temperature of $\lesssim 1,800$ K. We constrain the donor mass with the period--density relationship for Roche-lobe-filling stars and find that $M_\textrm{donor} \lesssim 0.04\ M_\odot$. The binary parameters are consistent with evolutionary models for post-period minimum CVs, suggesting that SRGeJ0411 is a new period bouncer. The optical emission lines of SRGeJ0411 are single-peaked despite the system being eclipsing, which is typically only seen due to stream-fed accretion in polars. X-ray spectroscopy hints that the white dwarf in SRGeJ0411 could be magnetic, but verifying the magnetic nature of SRGeJ0411 requires further investigation. The lack of optical outbursts has made SRGeJ0411 elusive in previous surveys, and joint X-ray and optical surveys highlight the potential for discovering similar systems in the near future.

\end{abstract}




\section{Introduction}

Cataclysmic variables (CVs) are compact object binaries in which a white dwarf (WD) accretes from a Roche-lobe filling donor, typically a late-type main-sequence star. The donor fills its Roche lobe and develops a teardrop-like shape, with the tip positioned at the Lagrangian $\rm L_1$ point. Matter leaves the secondary star through the vicinity of this point and forms an accretion stream. When the Alfvén radius of the WD does not extend past its surface, the WD accretes material from the donor via an accretion disk \citep[e.g.][]{warner95, hellierbook}. In the case of a strong magnetic field, the accretion disk can be truncated (intermediate polars; $B\approx 1-10$ MG) or may even be prevented from forming (polars; $B\approx 10-250$ MG). In these cases, the accreted material is channelled along magnetic field lines before hitting the surface of the WD.

In the canonical picture, these systems are formed through common envelope evolution \citep{1976IAUS...73...75P}. Angular momentum loss (AML) of the binary system then drives it to shorter orbital periods as the secondary loses mass to WD and the system as a whole loses matter and orbital angular momentum. At orbital periods above $\approx$ 2--3 hr, magnetic braking is the dominant contributor to AML. Below $\approx$ 2 hr, mass loss by the secondary gradually extinguishes nuclear burning, and the star becomes a low-mass degenerate object. The mass-radius relation index becomes negative (i.e. the radius increases as mass is lost), mean density of the donor decreases and the system begins to evolve to longer periods (see Eq.~(\ref{eq:pho_donor}) below). Hence, there should be a period minimum \citep{1971faulkner, 1981paczynski}. Systems that have undergone this transition are called "period bouncers".
The observed value of this period is around 78 min.\footnote{This period minimum is the approximate value for CVs with late-type donors that fill their Roche lobes while still on the main sequence (this constitutes the majority of CVs). The "period minimum" can be much lower for "evolved CVs" --- the systems where the donor overfilled its Roche lobe at the verge of hydrogen exhaustion in its core ($X_c \lesssim 0.1$) or even having a miniscule ($M \sim 0.01\,M_\odot$) helium core 
\citep{1985SvAL...11...52T,1987SvAL...13..328T,2003MNRAS.340.1214P}}.

Population synthesis studies predict that around $\approx$ 40--75\% of Galactic CVs are post-period minimum systems \citep[see, e.g.,][]{1993A&A...271..149K, 2015ApJ...809...80G,2020belloni}. 
This is in a sharp contrast with observational data. Using a volume-limited ($\leq$150 pc) sample of 42 CV from different sources with accurate parallaxes from {\it Gaia} DR2, \citet{2020pala} estimated a space density of CV to be $(4.8^{+0.6}_{-0.8})\times 10^{-6}\,\mathrm {pc^{-3}}$ and the space density of bouncers to be $\approx 0.3 \times 10^{-6}\,\mathrm {pc^{-3}}$. \citet{2023MNRAS.525.3597I} analysed spectroscopy for the sample of 118 CVs from SDSS and found a space density of CVs equal to  $7.8 \times 10^{-6}\,\mathrm {pc^{-3}}$, out of which  $\simeq(0.2 \times 10^{-6}\,\mathrm{ pc^{-3}})$ are bouncers. Thus, the period bouncers comprise only about (3 -- 7)\% of observed CVs \citep[see ][for compiled samples of confirmed and candidate objects]{2011MNRAS.411.2695P,2018PASJ...70...47K, 2021ApJ...918...58A,2023MNRAS.525.3597I}.

The donors in period bouncers are brown dwarfs (L/T spectral types), which allow for the WD and/or accretion disk to dominate in the optical \citep{2011knigge}. Moreover, the intrinsic optical faintness of period bouncers, along with their low accretion rates, which lead to infrequent dwarf nova outbursts, makes them challenging to detect. It is even more difficult to precisely determine the donor mass and therefore confirm these systems as period bouncers, often requiring near-infrared spectroscopy \citep{2006Sci...314.1578L}. 

Period bouncers may be hidden among Tremendous Outburst Amplitude dwarf Novae, like V592~Her, a system that harbours a brown dwarf \citep{1999A&A...342L..45V,1998ASPC..137..207P,2002A&A...395..557M}. Other (speculative) suggested solutions of the problem include merger of components due to unstable mass-loss after the "last" nova eruption in the system \citep[e.g.][]{2016ApJ...817...69N} or contraction of brown dwarf donor and cessation of mass-loss. Even more extreme version of the latter suggestion is that of \citet{1998PASP..110.1132P} on existence of a permanent source of AML, additional to gravitational wave radiation, that speeds up evolution of bouncers and their transformation into planets. \citet{Schreiber2023} speculated that spin-orbit synchronisation effects due to the action of rotation- and crystallization-driven dynamo effects may cause detachment of post-period-minimum mass donors from the Roche lobe and termination of mass transfer.

While most period bouncers are non-magnetic CVs, several candidate-period bouncers are spectroscopically confirmed to harbour a magnetic WD \citep[e.g. SDSSJ121209.31+013627.7 (V379 Vir), SDSSJ151415.65+074446.5, SDSSJ125044.42+154957.4][]{2006MNRAS.373.1416B,2008ApJ...674..421F, 2009kulebi,2012MNRAS.423.1437B}. Observational data, including X-ray observations, is not sufficient to distinguish whether these systems harbour a Roche-lobe filling component or a detached one \citep[see][]{2012MNRAS.423.1437B,2017A&A...598L...6S,2023A&A...676A...7M}. While in a 150 pc volume-limited sample of CVs, it was found that 36\% of CVs are magnetic, the fraction of magnetic period bouncers remains uncertain \citep{2020pala}.

All-sky surveys provide possibilities to study the Galactic CV population. The eROSITA telescope aboard the Spektr-RG mission \textrm{\citep[SRG;][]{2021sunyaev, 2021erosita}} goes $\sim$10 times deeper than the previous all-sky Roentgensatellit X-ray survey \citep[ROSAT;][]{1982rosat1}, with improved localisation of X-ray sources, and has already led to the discovery of new CVs \citep[e.g.][]{2022schwope, 2022AstL...48..530B, 2023rodriguez}. Some CVs were discovered using the Mikhail Pavlinsky ART-XC telescope \citep{2011SPIE.8147E..06P} on board of SRG observatory \citep[e.g.][]{2022A&A...661A..39Z}. In addition, some well-known period bouncers were studied with SRG/eROSITA \citep{2021A&A...646A.181S,2023A&A...676A...7M}.

A multi-wavelength campaign, combining X-ray and optical information, gives the possibility to search for and discover CVs. SRGeJ0411 is one of the objects identified as a CV candidate in a cross-match of a $\rm 1200\ deg^2$ patch of the sky of SRG/eROSITA X-ray data and the optical ZTF database. SRGeJ0411 was called to our attention by its high ratio of X-ray flux to optical flux, $\rm F_{X}/F_{opt}\approx 0.60$, and placement in the Gaia color-magnitude diagram near the WD region. A new 55 minutes period eclipsing AM CVn, SRGeJ045359.9+622444, was recently discovered from this joint SRG/eROSITA and ZTF search \citep[][hereafter \citetalias{2023ApJ...954...63R}]{2023ApJ...954...63R}. More targets identified from this program will be presented in future work (Galiullin et al. in prep.).

In this paper, we present the study and characterisation of the new eclipsing CV with evidence of a brown dwarf secondary, SRGeJ0411. In Section \ref{sec:data}, we briefly describe the  X-ray and optical observations. The data reduction was similar to the ones used in \citetalias{2023ApJ...954...63R}. In Section \ref{sec:results}, we analyse the optical high-speed photometry, phase-resolved spectroscopy, spectral energy distribution (SED), and X-ray spectroscopy of SRGeJ0411 in order to solve for the binary parameters. We also  place SRGeJ0411 on CV evolutionary tracks. In Section \ref{sec:discussion}, we discuss the possible magnetic nature of SRGeJ0411. We summarise our results in Section \ref{sec:summary}.

\section{Data and Observations}
\label{sec:data}

\begin{figure*}
\begin{center}
\includegraphics[width=0.83\textwidth]{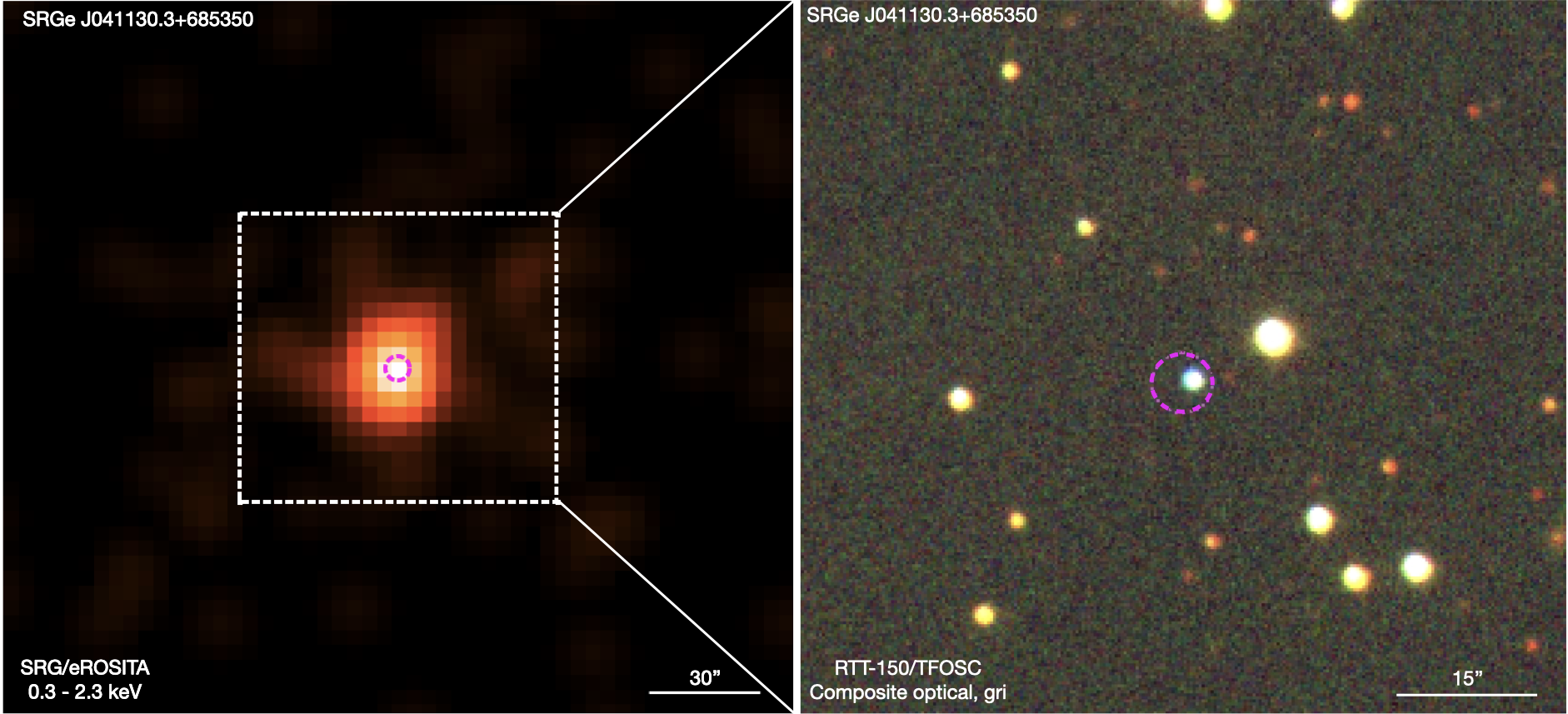}
\caption[] { {\it Left:} False-colour X-ray image of SRGeJ0411 in the 0.3--2.3 keV energy band from combined data of four all-sky surveys of SRG/eROSITA. The image was smoothed with a $15\arcsec$ Gaussian kernel. {\it The white box} shows the field of view of the optical image on the right. {\it Right:} Composite optical image around SRGeJ0411 based on RTT-150/TFOSC data. A pseudo-colour image was composed using $gri$ filters. {\it The magenta circle} with a radius of  3.3$\arcsec$ (98\% localisation error, R98) is centred at the X-ray position of SRGeJ0411.}
\label{fig:images}
\end{center}
\end{figure*}

\begin{figure}
    \centering
    \includegraphics[width=0.45\textwidth]{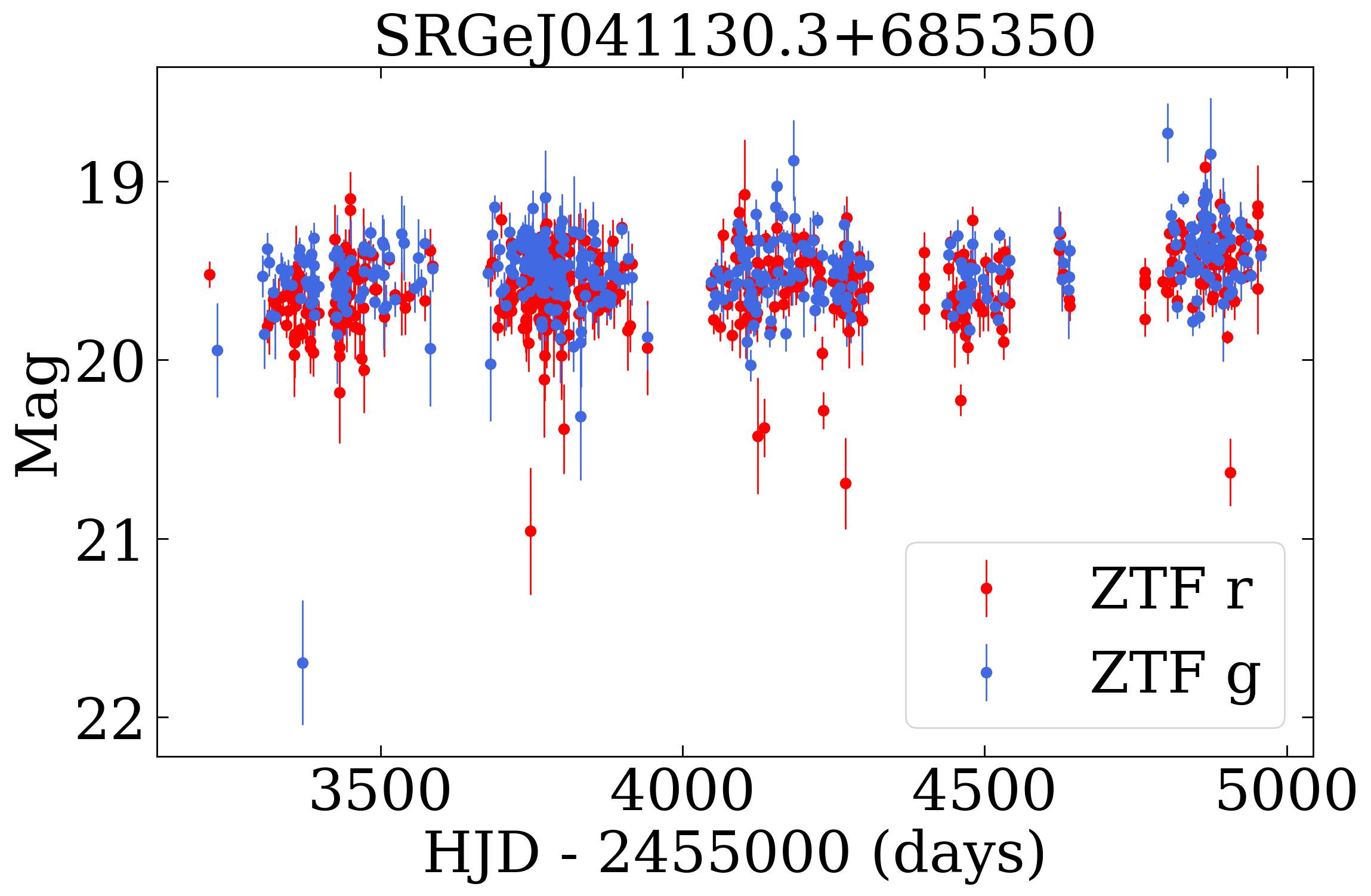}\\
    \includegraphics[width=0.43\textwidth]{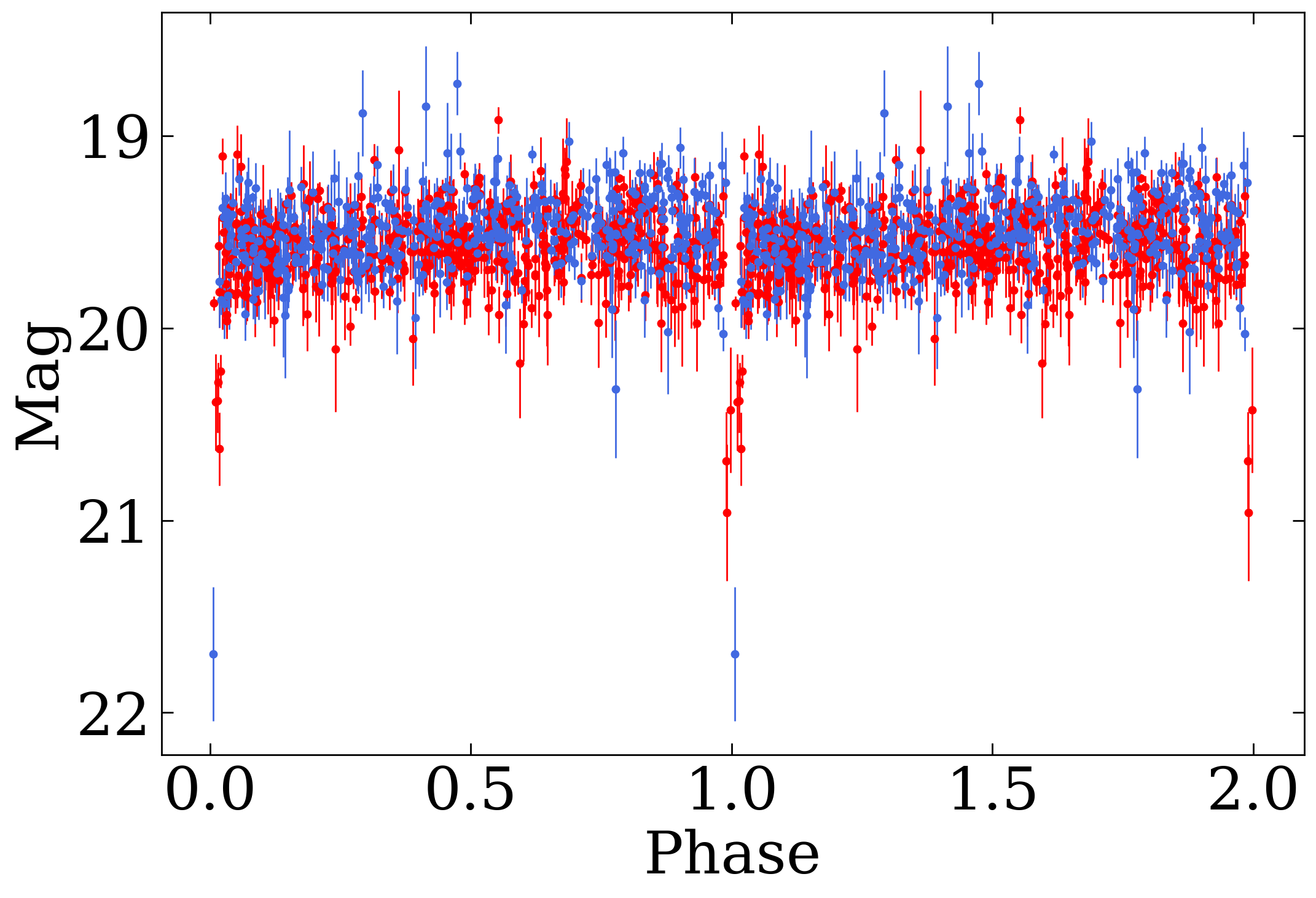}
    \caption{ZTF light curves of SRGeJ0411 in $g$ and $r$ filters: long-term ({\it top}), and folded at the 97.5 minutes orbital period ({\it bottom}). No significant outbursts are seen over the $\approx$5 yr-long baseline.}
    \label{fig:ztf_lc}
\end{figure}

\begin{figure}
    \centering
    \includegraphics[width=0.45\textwidth]{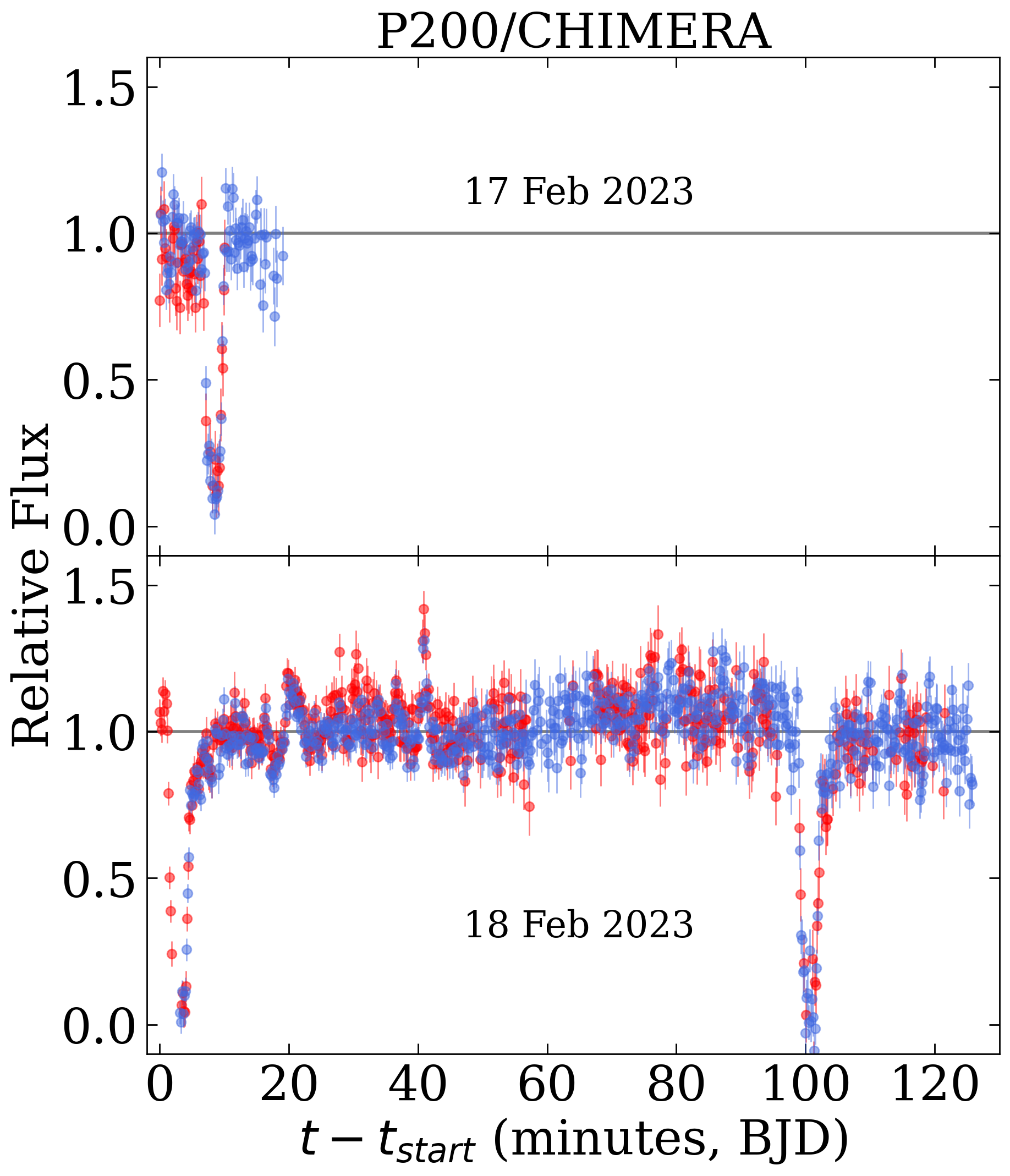}
    \includegraphics[width=0.5\textwidth]{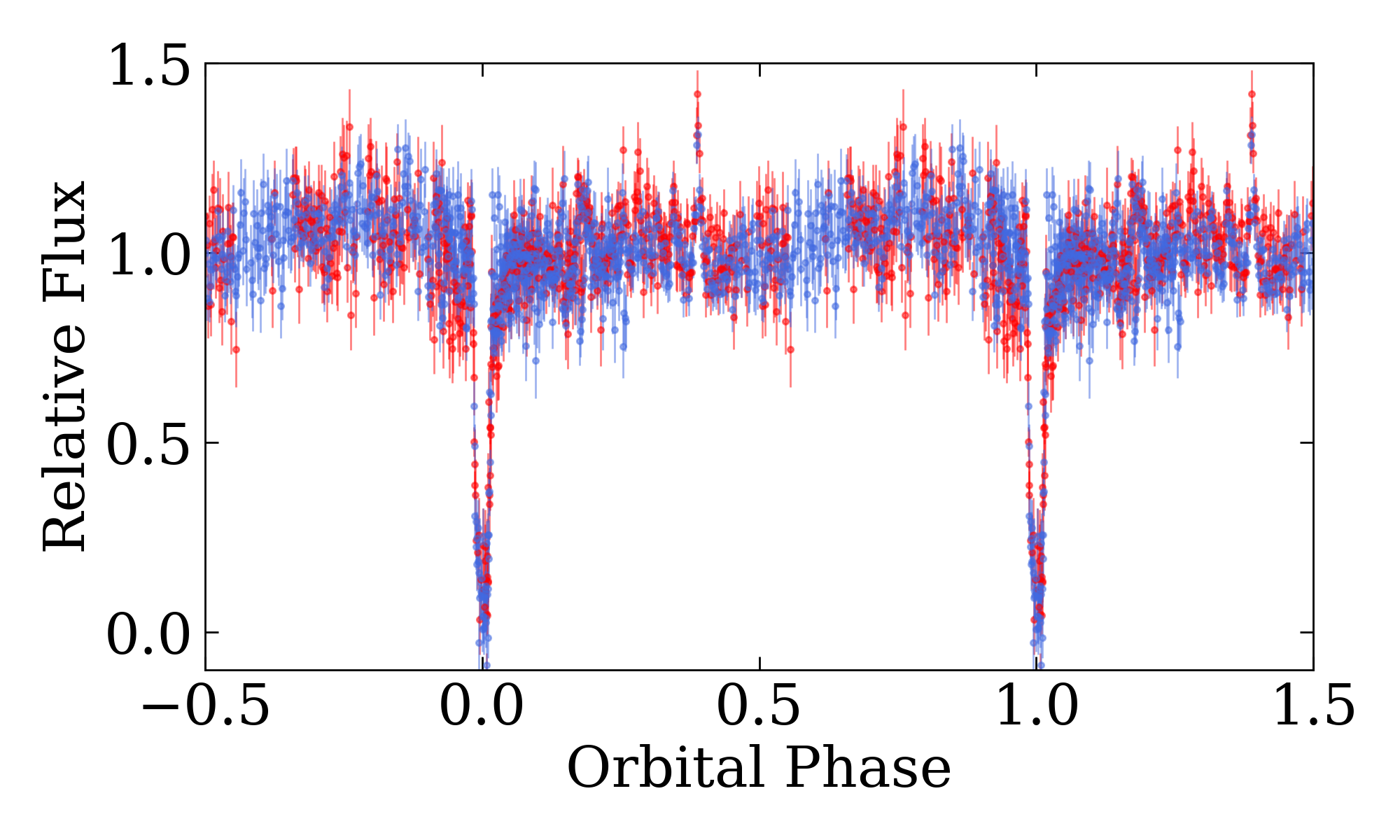}
    \caption{CHIMERA $r$ ({\it red}) and $g$ ({\it blue}) 10-sec cadence photometry reveal deep eclipses. \textit{Upper two panels}: The entire observation on each occasion. Gaps are due to large error bars in the data, where cloud cover or highly variable seeing prevented a good extraction of the data. \textit{Bottom panel}: Data from 17 and 18 February 2023 are folded over the 97.53-minute orbital period.}
    \label{fig:CHIMERA}
\end{figure}

The data reduction was done following the standard techniques and procedures, similar to the ones described in Section 2 of \citetalias{2023ApJ...954...63R}. Figure \ref{fig:images} (left panel) shows the X-ray image of the SRGeJ0411 obtained from combined data from four all-sky surveys of the SRG/eROSITA. We present archival ZTF data (SNR $>$ 5; no upper limits shown) for SRGeJ0411 in Figure \ref{fig:ztf_lc}. Table \ref{tab:data}  summarises the timeline of all observations,  wavelength coverage and resolution information. Here, we briefly summarise the data and observations used in this work at the difference to \citetalias{2023ApJ...954...63R}:

\begin{itemize}

\item[--] {\it LRIS spectroscopy}: We obtained an identification spectrum of SRGeJ0411 on the Keck I telescope using the Low-Resolution Imaging Spectrometer \citep[LRIS;][]{lris} on 15 January 2023 (UT). We used a 1.0$\arcsec$ slit, and the seeing during the portion of the night was approximately 0.7$\arcsec$, leading to minimal slit losses. 

\item[--] {\it DBSP spectroscopy}: We obtained phase-resolved spectroscopy using the Double Spectrograph \citep[DBSP;][]{dbsp} on the Hale telescope on 27 March 2023. We used the 600/4000 grism on the blue side and the 316/7500 grating on the red side. A 1.0$\arcsec$ slit was used, and the seeing throughout the observation varied between 1.0 -- 1.2$\arcsec$, leading to some slit losses. That night, all exposures were forced to be taken at an elevation below 45 degrees (corresponding
to airmass $\approx$ 1.4) as the object was quickly setting. All P200/DBSP data were reduced with \texttt{DBSP-DRP}\footnote{\url{https://dbsp-drp.readthedocs.io/en/stable/index.html}}, a Python-based pipeline optimized for DBSP built on the more general \texttt{PypeIt} pipeline \citep{2020pypeit}. All data were flat fielded sky-subtracted using standard techniques. Internal arc lamps were used for the wavelength calibration and a standard star for overall flux calibration. 

\item[--] {\it CHIMERA high-speed photometry}: We acquired high-speed photometry in $r$ and $g$ bands using the Caltech HIgh-speed Multi-color camERA \citep[CHIMERA;][]{chimera} on two occasions. We are clearly able to identify eclipses in all CHIMERA data sets, but it may be difficult to interpret out of eclipse variability. Seeing was abnormally poor on each night: 3$\arcsec$ on 17 February 2023 and 18 February 2023. We show all individual light curves and the combined light curve in Figure \ref{fig:CHIMERA}.

\item[--] {\it RTT-150}: We performed SRGeJ0411 photometry with the 1.5-meter Russian-Turkish Telescope (RTT-150). The resolution element is 0.65$\arcsec$ at $2\times 2$ binning. Three sets of observations in the white filter with a duration of two hours per night were carried out on September 1, 6 and 8, 2023. The weather was clear, and the average seeing was 1.5--2.0$\arcsec$. The time resolution was 55 sec in all sets of observations (with exposure time of 30 sec and readout time of 25 sec). The photometry of SRGeJ0411 was calibrated to the \textit{Gaia} G band, taking into account the \textit{Gaia} BP-RP colour correction (systematic shift of $-0.25^m$ relative to nearby stars).  Figure \ref{fig:images} (right panel) shows the optical image near SRGeJ0411 from RTT-150 data. On September 11, we obtained additional 3 x 300 sec frames in griz filters at $1\times 1$ binning (resolution element is 0.325$\arcsec$) and used PanStarrs photometry to calibrate the field stars and determine griz magnitudes of SRGeJ0411 (see Figures \ref{fig:images}, \ref{fig:rtt_wise} and \ref{fig:colors_sed}).

\end{itemize}

\begin{table*}
\fontsize{8}{8}\selectfont
\centering
\caption{Data Acquired for SRGeJ0411.}
\label{tab:data}
\begin{tabular}{ccccc}
\hline
Data Type & Date (UT) & Instrument & Specifications & Finding\\
\hline
{\parbox{2cm}{\vspace{5pt}\centering Identification Spectrum}} & {\parbox{2cm}{\vspace{5pt}\centering 15 Jan. 2023}} &{\parbox{4cm}{\vspace{5pt}\centering Keck I/LRIS}}& {\parbox{4cm}{\centering \vspace{5pt}Blue: 3140--5640 \AA, \\$\Delta \lambda = $1.1 \AA, 1x900s exp.  \\ Red: 5530--8830 \AA,\\ $\Delta \lambda = $0.80 \AA, 1x900s exp.}}&{\parbox{4cm}{\vspace{5pt}\centering Single-peaked emission lines are discovered, but double-peaked lines are expected due to eclipse seen in ZTF.\vspace{0.2cm}}}\\\hline

{\parbox{2cm}{\centering \vspace{5pt}High-cadence $r$ and $g$ band Photometry}}
 & {\parbox{2cm}{\vspace{5pt}\centering 17, 18 Feb. 2023}}  & {\parbox{3cm}{\vspace{5pt}\centering Hale Telescope/CHIMERA}} &{\parbox{4cm}{\centering \vspace{5pt}10s exp. for 2 hr}} & {\parbox{4cm}{\vspace{5pt}\centering High-cadence photometry at simultaneous orbital phases confirms deep eclipse.}}\\ \hline
 
{\parbox{2cm}{\centering \vspace{5pt}Multi-phase spectra}} & {\parbox{2cm}{\vspace{5pt}\centering 27 Mar. 2023}} &{\parbox{2cm}{\vspace{5pt}\centering Hale Telescope/DBSP}}& {\parbox{4cm}{\centering \vspace{5pt}Blue: 3400--5600 \AA, \\$\Delta \lambda = $1.5 \AA, 7x900s exp.  \\ Red: 5650--10,200 \AA,\\ $\Delta \lambda = $1.1 \AA, 7x900s exp.}}&{\parbox{4cm}{\vspace{5pt}\centering Single-peaked emission seen at all orbital phases.}} \\ \hline

{\parbox{2cm}{\centering \vspace{5pt} Photometry, no filter}}
 & 1, 6, 8 Sep. 2023 & {\parbox{4cm}{\vspace{5pt}\centering RTT-150/TFOSC }}&{\parbox{4cm}{\centering \vspace{5pt} 30s exp. for 2 hrs}} & {\parbox{4cm}{\vspace{5pt}\centering Precision of period improved; Photometry revealed flickering and non-eclipse magnitude change between observations.} }

\\
\hline
\end{tabular}

\end{table*}

\begin{figure}
    \centering
    \includegraphics[width=0.45\textwidth]{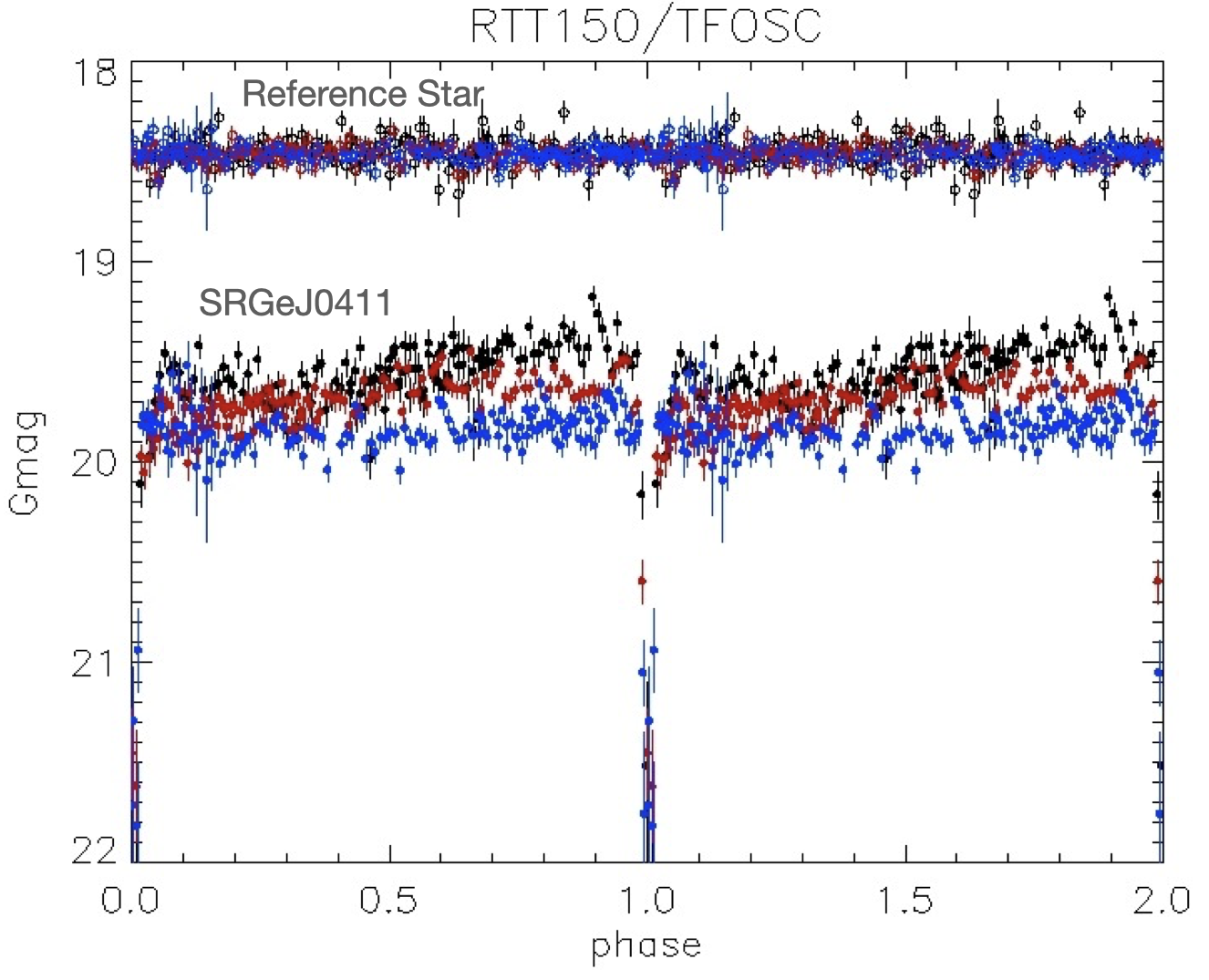}    
    \caption{Phase-folded light curve of RTT-150 data. {\it Color dots} correspond to different observation dates: {\it black} -- September 1; {\it red} -- September 6; {\it blue} -- September 8, 2023. Parts of light curves out of the eclipse are not constant, showing a magnitude variation, possibly caused by variation of the accretion rate in SRGeJ0411. The reference star magnitude is offset by $\rm 1^m$ for illustrative purposes.}
    \label{fig:rtt_lc}
\end{figure}

\section{Results}
\label{sec:results}

\subsection{ The Light Curve and the Orbital Period}
\label{sec:period}

\begin{figure}
\centering
\includegraphics[width=0.45\textwidth,clip=true]{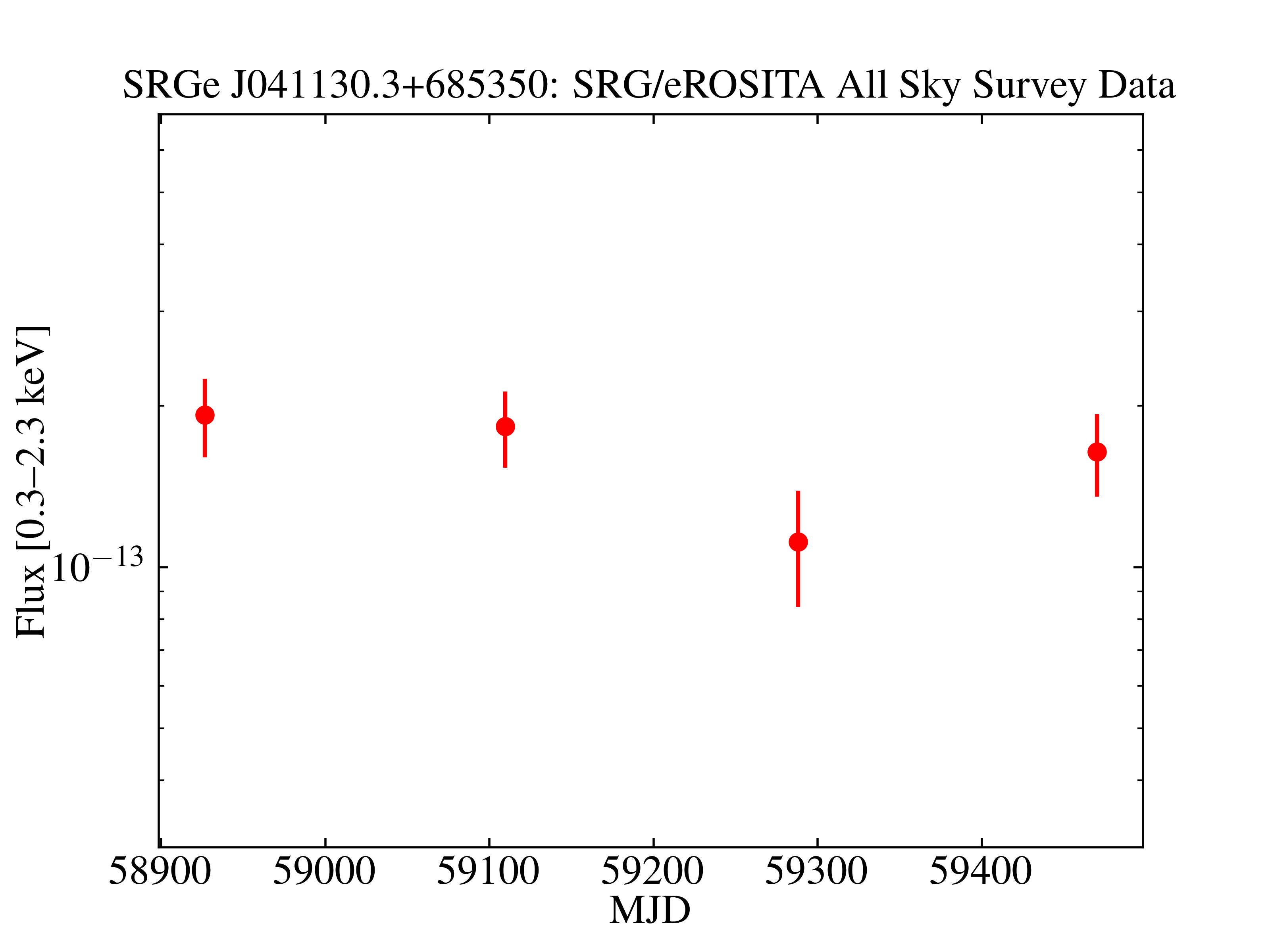}
\caption{The 0.3--2.3 keV X-ray light curve of SRGeJ0411 during four SRG/eROSITA all-sky surveys.}
\label{fig:Xray_LC}
\end{figure}

The optical light curve of SRGeJ0411 shows deep eclipses ($\approx2.5^m$) and no significant outbursts over the five-year-long baseline on ZTF data (see Figure \ref{fig:ztf_lc}). We used a technique based on the Box Least Squares (BLS) algorithm to determine the orbital period of SRGeJ0411 as described in \citetalias{2023ApJ...954...63R}. We found the best-fit period of $97.5\pm0.5$ minutes using the ZTF forced photometry data. The ZTF optical light curve of SRGeJ0411 folded with the best-fit period is shown in Figure \ref{fig:ztf_lc} (lower panel).

Following \citetalias{2023ApJ...954...63R}, we better constrained the orbital period of SRGeJ0411 using CHIMERA observations. 
We computed the  $97.530 \pm 0.008$ minutes orbital period and ephemeris $t_0$(BJD)= 2459992.6769(5). We present all good quality CHIMERA data folded on this period in Figure \ref{fig:CHIMERA}.

We searched for the period of SRGeJ0411 on RTT-150 data using frequency analysis and the Lafler-Kinman method (for more details see Appendix A in \citetalias{2023ApJ...954...63R}). The periodogram gives a strongest period of $97.53552\pm0.00936$ minutes, which agrees with the CHIMERA period. Having an initial epoch $t_0$ based on the Chimera observations and a time difference of 6.5 months between CHIMERA and RTT-150 observations, we can improve the precision of the orbital period of SRGeJ0411. Thanks to the combination of both RTT-150 and CHIMERA data, we found the improved orbital period to be $P_\textrm{orb} = 97.529544\pm0.000173$ minutes by dividing the resulting phase difference of the RTT-150 and CHIMERA light curve by the number of cycles ($\sim$2900 cycles) that have passed between observations. In all analyses, we adopted the orbital period of $P_\textrm{orb}\approx$ 97.530 minutes, and ephemeris $t_0$(BJD)= 2459992.6769(5).

The RTT-150 phase-folded optical light curves of SRGeJ0411 and the reference star for three nights of observations are shown in Figure \ref{fig:rtt_lc}. The phase-folded light curve of the reference star\footnote{The celestial coordinate for the reference star used in the RTT photometry is RA(J2000.0)=$04^{h} 11^{m} 34.83^{s}.1$ and DEC(J2000.0)=$+68\degr 53^{'} 48^{''}.0$, with a {\it Gaia} G band magnitude of $\rm 19.45^m$.} stays constant between observations, having a root-mean-square of $0.05^m$. All optical light curves show low amplitude ($\approx0.2^m-0.5^m$) flickering. Between observing nights, the light curves out of the eclipse show a magnitude change with an amplitude of about $0.5^m$. The light curve change of SRGeJ0411 within observations is not an instrumental effect and could be caused by variations of accretion rate in the system.  We cannot provide additional information about the variability between different nights using CHIMERA data because only in one night (18 February 2023) was an entire orbital period of SRGeJ0411 covered.
  
The CHIMERA light curves show a $\approx$20\% modulation in flux through excess brightness before the eclipse (see Figure \ref{fig:CHIMERA}), possibly caused by the contribution of the bright spot emission and/or eclipse of an accretion stream (see Section \ref{sec:discussion}).  Conversely, the X-ray light curve based on four SRG/eROSITA all-sky survey data shows no significant variability. The $\chi^2$/dof is 5.7/3, assuming the constant X-ray flux of the SRGeJ0411 within surveys (see Figure \ref{fig:Xray_LC}). SRGeJ0411 requires further X-ray follow-up to get more data on a light curve and verify possible X-ray variability.

\subsection{Optical Spectroscopy}

\begin{figure*}
    \centering
    \includegraphics[width=\textwidth]{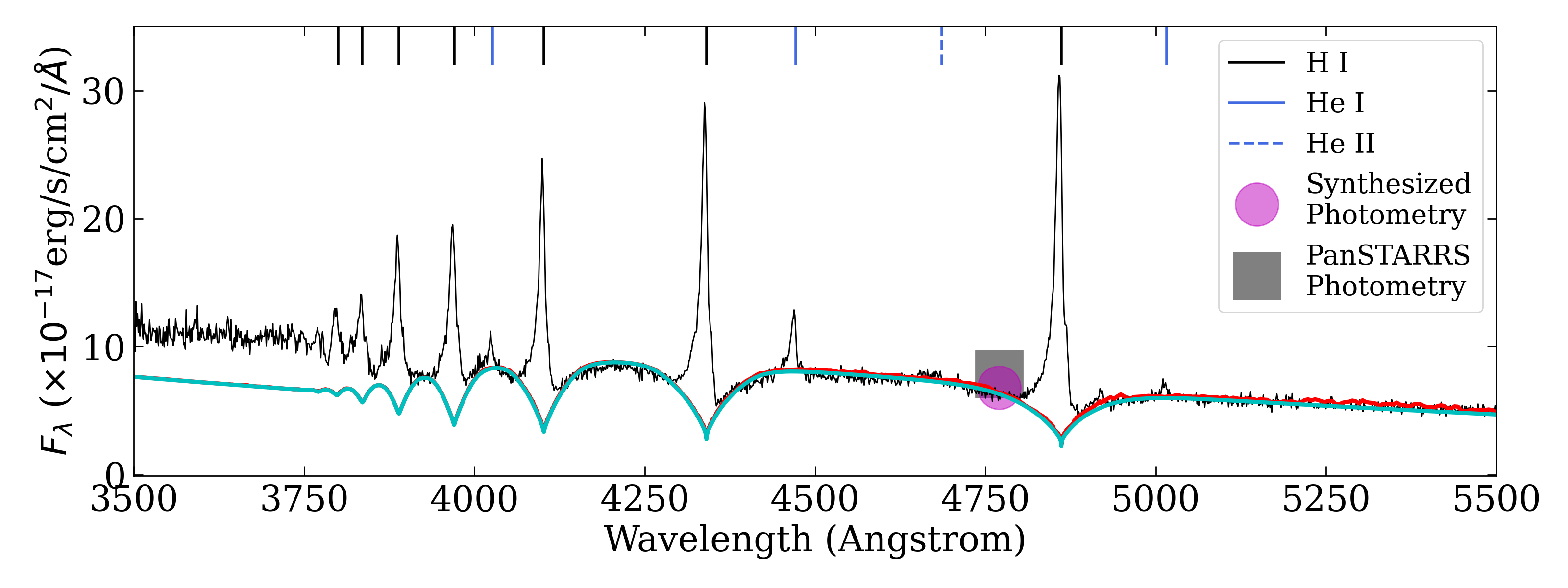}\\
    \includegraphics[width=\textwidth]{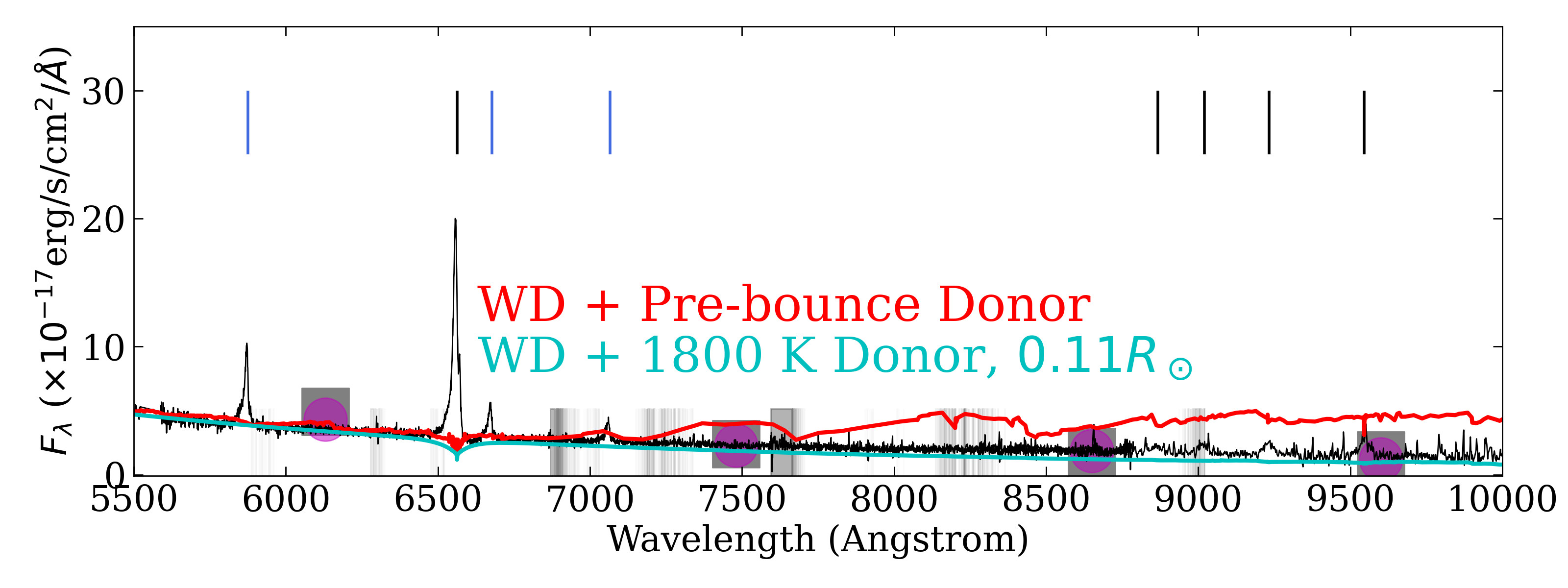}
    \caption{ Keck I/LRIS optical spectrum of SRGeJ0411. {\it Grey lines} are locations where there are telluric features from the Keck Telluric Line List. {\it Grey boxes} indicate Pan-STARRS photometry, which differ from the synthesized photometry ({\it magneta circles}) to within 10\%, likely due to instrinsic source variability. In {\it red}, we show the model spectrum of a WD + a 2,900 K donor star if the system were a pre-bounce CV at $P_\textrm{orb}=97.5$ minutes at the distance of SRGeJ0411. Since a disk and/or a bright spot would only contribute more flux, this plot, in conjunction with the known period or SRGeJ0411, is sufficient to confirm its nature as a period bouncer (see Section \ref{sec:keck_spectrum} for more details). In {\it cyan}, we show the model spectrum of a WD + a donor star ($T_\textrm{eff, donor} \approx 1,800$ K, $R_\textrm{donor}=0.11\ R_\odot$), where donor parameters are computed from SED modelling (see Sections \ref{sec:sed}). For both model spectra ({\it red} and {\it cyan}), WD parameters are also computed from SED modelling.}
    \label{fig:LRIS}
\end{figure*}

\begin{table}
	\centering
	\caption{Equivalent widths (EWs) of selected lines.
	}
	\label{tab:EW_lines}
	\begin{tabular}{lc} 
		\hline
		Line (\AA) &   {\parbox{2cm}{\vspace{5pt}\centering DBSP 27 Mar. EW (\AA)}} \\
		\hline
\textit{Emission features} &\\
H$\alpha$\ 6562.8 & $-112.5\pm 0.5$\\
H$\beta$\ 4861.3 & $-47.7\pm 0.5$\\
H$\gamma$\ 4340.5 & $-17.5\pm 0.6$\\
H$\delta$\ 4101.7 & $-2.5\pm 0.4$\\
H$\epsilon$\ 3970.1 & $-2.1\pm 0.6$\\
He I\ 4471.4  & $-4.8\pm 0.2$\\
He I\ 5876.5  & $-19.8\pm 0.4$\\
He I\ 6678.2 & $-7.1\pm 0.6$\\
He I\ 7065.2  & $-4.6\pm 0.7$\\
He I\ 7281.4  & $-4.5\pm 0.6$\\
He II\ 4685.7  & $-1.4\pm 0.7$\\
\hline
	\end{tabular}
\end{table}

\subsubsection{Identification Spectrum}
\label{sec:keck_spectrum}

In Figure \ref{fig:LRIS}, we present the Keck I/LRIS identification spectrum of SRGeJ0411. The optical spectrum shows hydrogen and helium emission lines, which are seen in almost all CVs. We see only single-peaked emission lines and do not detect double-peaked lines as expected from the accretion disk of eclipsing non-magnetic CVs \citep[e.g.][]{1980cv_disk_lines, warner95}. The absence of line doubling in the optical spectrum of SRGeJ0411 is discussed in Section \ref{sec:discussion}. The Balmer jump in emission is also seen in Figure \ref{fig:LRIS}, which is seen due to accretion in both magnetic and non-magnetic CVs \citep[e.g.][]{hellierbook}. We chose to present the Keck I/LRIS spectrum since it was obtained at a more favourable airmass and has a higher signal-to-noise ratio than the phase-averaged P200/DBSP spectrum. In Figure \ref{fig:LRIS}, we overplot both the Pan-STARRS \citep{2016panstarrs} photometry and photometry synthesized from the Keck spectrum to the Pan-STARRS bandpasses using the \texttt{pyphot} package\footnote{\url{https://mfouesneau.github.io/pyphot/index.html}}. All fluxes agree to within 10\%, likely due to a combination of the Keck spectrum having been taken at a single orbital phase as well as intrinsic source variability. However, we use the P200/DBSP spectra to conduct a detailed analysis of all lines, as outlined in the following subsection.

In Figure \ref{fig:LRIS}, we present two models: a WD + pre-bounce donor and a WD + 1,800 K donor alongside the data. We use the same WD parameters in both models, determined from the SED fit to UV + optical data ($T_\textrm{eff, WD}$ = 13,780 K, $R_\textrm{WD}=0.01\ R_\odot$, log(g) = 8.0; see Section \ref{sec:sed}). We then use the CV evolutionary tracks from \cite{2011knigge} to determine the pre-bounce donor parameters at a 97.5-minute orbital period: $T_\textrm{eff, donor} = 2,900\ K$, $R_\textrm{donor} = 0.15\ R_\odot$, log(g) = 5.0. Both the ``standard'' and ``optimal''\footnote{ \citet{2011knigge} suggest that agreement between evolutionary theory and observations may be improved if ``standard'' expressions for angular momentum loss via magnetic braking above the period gap \citep{1983ApJ...275..713R} and via gravitational waves radiation below the gap \citep{1967AcA....17..287P} would be scaled by  $f_{\rm MB} = 0.66 \pm 0.05$ and $f_{\rm GR} = 2.47 \pm 0.22$ respectively. But note, physical justification for such a scaling is absent.}  (revised) tracks give close values for a pre-bounce donor, but we adopt the ones from the ``optimal" track since it still slightly better reproduces observed CVs. We assume a solar metallicity and plot a BT-DUSTY model spectrum \citep{2011btdusty}, smoothed and binned to the average resolution of LRIS in Figure \ref{fig:LRIS} summed with the WD (red colour). We show (in cyan colour) the possible contribution of a donor ($T_\textrm{eff, donor} \approx 1,800$ K, $R_\textrm{donor}=0.11\ R_\odot$) and WD (same parameters as above), where donor parameters are computed from SED modelling (see Section \ref{sec:sed}).

In Figure \ref{fig:LRIS}, it is clear that the spectrum of SRGeJ0411 does not present features typical of a pre-bounce CV. In other words, even by omitting the contribution of an accretion disk, stream, or bright spot, the observed spectrum of SRGe0411 does not agree with that of a WD + pre-bounce M dwarf according to the evolutionary tracks of \cite{2011knigge}. We also  discard the possibility of SRGeJ0411 being a pre-bounce "evolved CV" system. Systems like this, before their period bounce, have donors that dominate the optical spectrum and are hotter than M dwarfs \citep{2021pre-elms}.

\subsubsection{Phase-Averaged Spectrum}
\label{sec:optical_spectrum}

We analysed the average of all spectra taken on 27 March 2023 with P200/DBSP, which cover an entire orbit (with approximately a 10 percent gap due to read-out time). Table \ref{tab:EW_lines} shows equivalent widths (EWs) for prominent lines identified in the optical spectrum of SRGeJ0411, which we calculate from the averaged spectrum. 

We do not detect high-ionization lines such as He II 4685.7 \AA\  and the C III/N III Bowen blend at 4650 \AA, which are thought to originate in the accretion column of magnetic CVs \citep[e.g.,][]{2017AJ....153..144O}. The upper limit ($3\sigma$) for the EW ratio $\rm He II/H_{\beta}$ is 0.04. This suggests that based on this criterion alone \citep{1992silber}, SRGeJ0411 is not a CV with a strong magnetic field (i.e. a polar).  However, the possibility to be an intermediate polar may be discussed (see Section \ref{sec:discussion} for more details). We also do not detect any emission or absorption lines from a donor star or disk typically seen in CVs such as the Ca II triplet (8498, 8542, 8662  \AA), the Na I doublet (8183, 8195  \AA), or any TiO headbands seen in pre-bounce donors \citep[e.g.][]{hellierbook, 2002szkody}. This lack of pre-bounce donor features (or any donor features whatsoever), first suggested to us that SRGeJ0411 was a period bouncer with a brown dwarf donor. However, we also do not detect (beyond 1$\sigma$) any Si I 3906 \AA\; or Fe I triplet (5270, 5328, 5371 \AA) lines originating in the irradiated face of the donor as seen in the well-studied period bouncer BW Scl \citep[see Table 3 in][]{2023bw_scl}. However, in the Keck I/LRIS spectrum, there is possible evidence for the irradiated donor through a narrow component of H$\alpha$ in addition to the main broad emission line (see Section \ref{sec:discussion} and Appendix \ref{appendix:inv_dop}).

\begin{table}
\fontsize{8}{11}\selectfont
	\centering
	\caption{Radial Velocity Measurements of H  Lines.
	}
	\label{tab:rv_disk}
	\begin{tabular}{lcccc} 
		\hline
		Line (\AA) &   {\parbox{1cm}{\vspace{5pt}\centering $\gamma$ \\(km/s)}} & {\parbox{1cm}{\vspace{5pt}\centering $K_x$ (km/s)}}  &{\parbox{1cm}{\vspace{5pt}\centering $K_y$ (km/s)}} &{\parbox{1cm}{\vspace{4pt}\centering $K = \sqrt{K_x^2 + K_y^2}$ (km/s)}}\\
		\hline
H$\alpha$\ 6562.8 & $28\pm 3$& $-63\pm 6$& $216\pm 6$ & $225\pm 8$ \\
H$\beta$\ 4861.3  & $118\pm 5$& $-109\pm 6$& $235\pm 6$& $259\pm 8$\\
\hline

	\end{tabular}
\end{table}

\begin{figure*}
    \centering
    \includegraphics[width=0.6\textwidth]{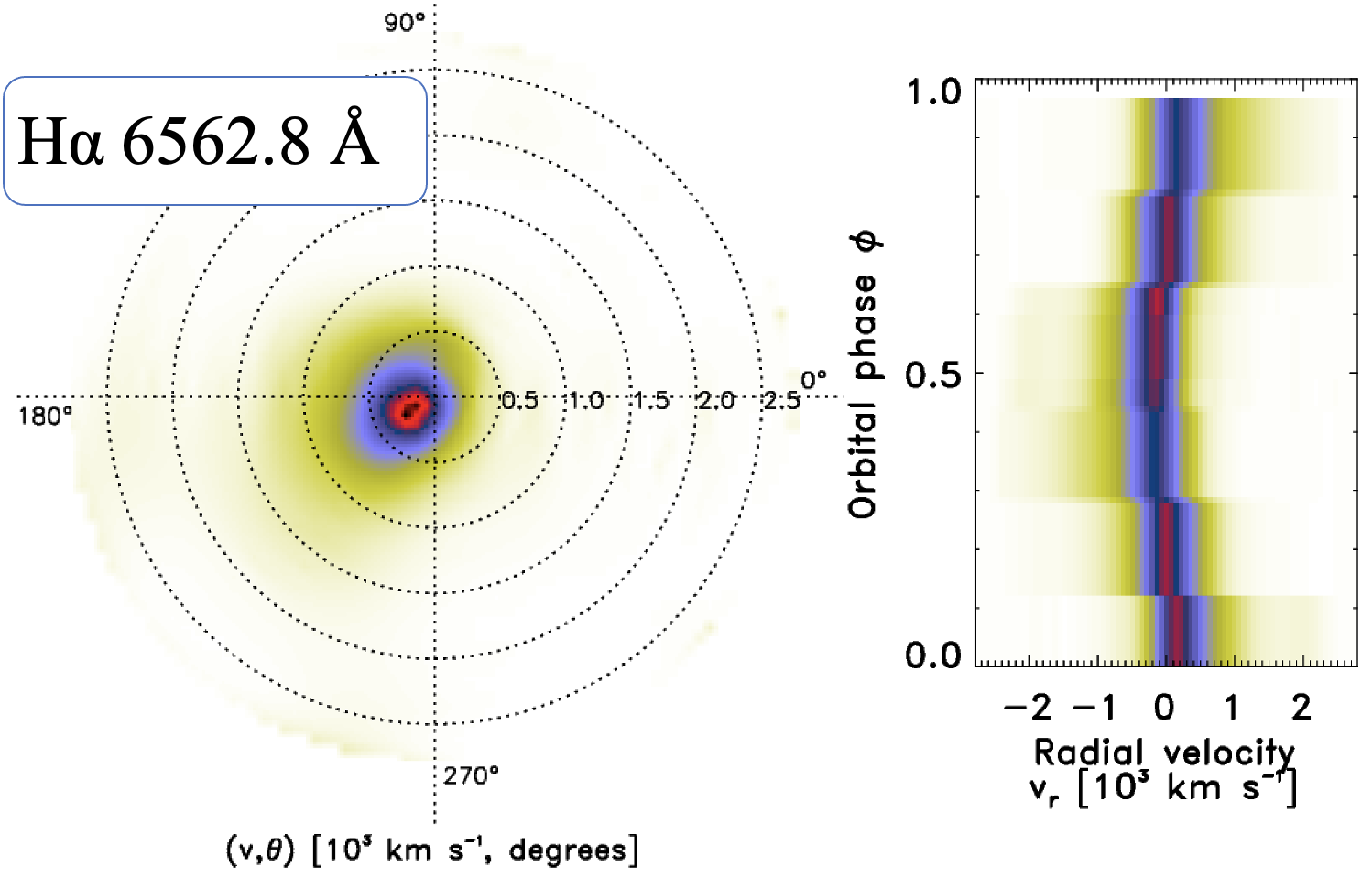} \includegraphics[width=0.2\textwidth]{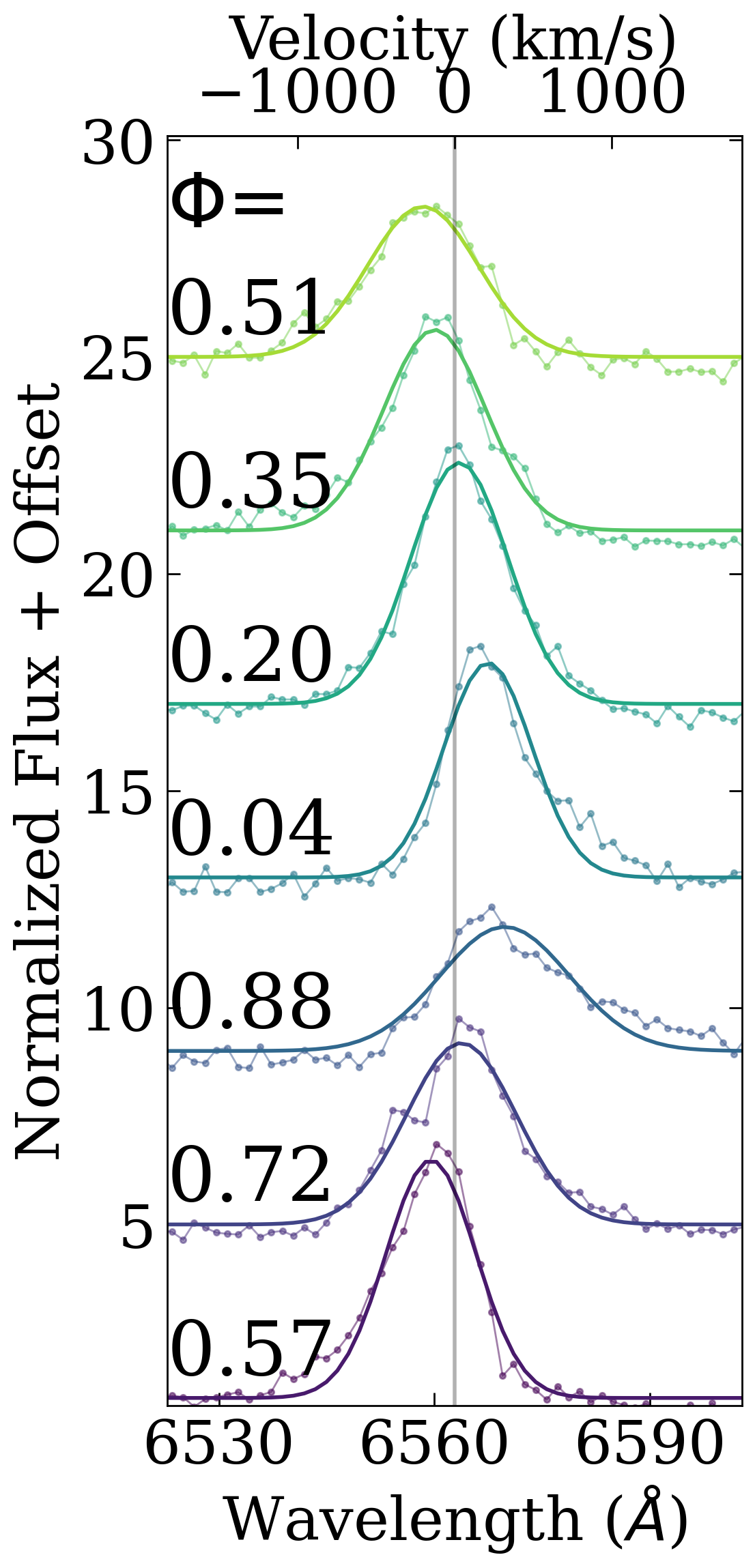}\\
    \includegraphics[width=0.6\textwidth]{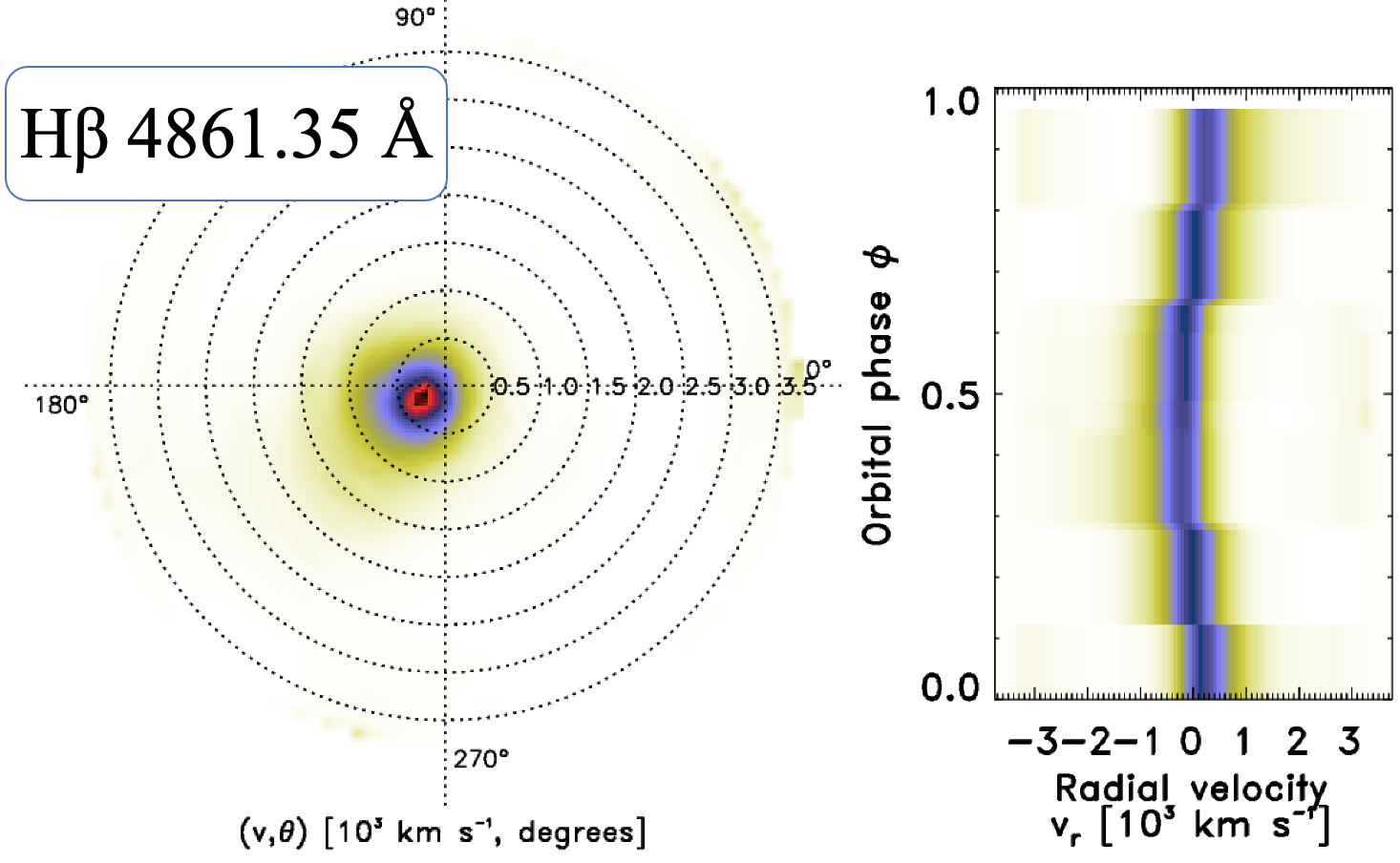}\includegraphics[width=0.2\textwidth]{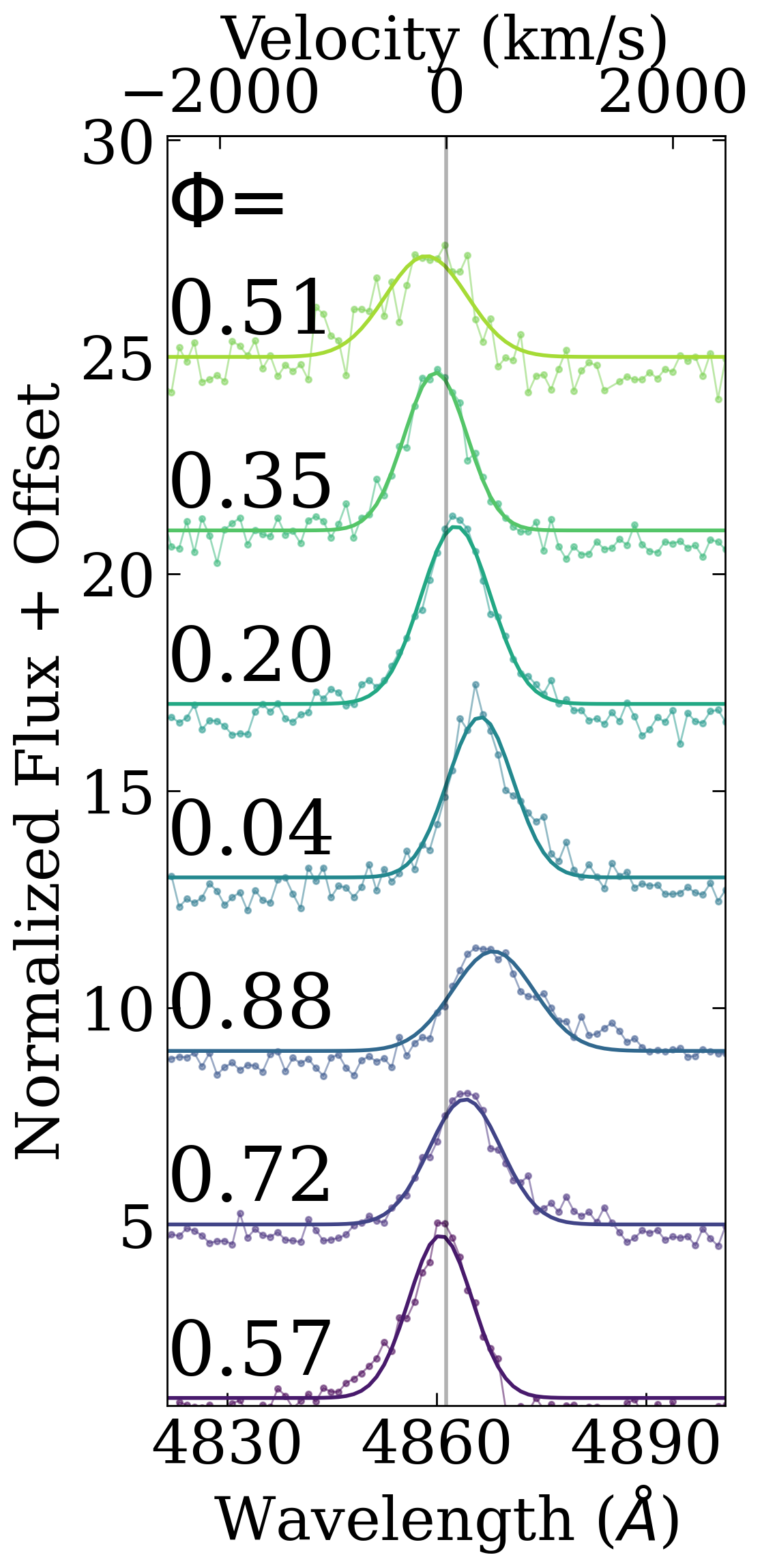}
    \caption{Doppler tomograms and trailed spectra of H$\alpha$ ({\it upper panel}) and H$\beta$ ({\it lower panel}) reveal a complete absence of an accretion disk. Instead, a broad, single-peaked emission line is seen throughout the orbit, with maximum radial velocity amplitudes when the WD and donor are aligned with respect to our line of sight. This points to either stream accretion or a bright spot (see Section \ref{sec:discussion} and Appendix \ref{appendix:inv_dop} for discussion).}
    \label{fig:doppler}
\end{figure*}

\subsubsection{ Phase-Resolved Spectra and Doppler Tomography}
For the $\rm H{\beta}$ 4861.35 \AA\ and $\rm H{\alpha}$ 6562.8 \AA\ emission lines, we construct Doppler tomograms and radial velocity (RV) curves, using seven DBSP spectra from a single orbit (for the details of the analysis see \citetalias{2023ApJ...954...63R}). Assuming a circular orbit, we have the following RV equation:
\begin{align}
    \textrm{RV} = \gamma +  K_x\sin(2\pi \phi) + K_y \cos(2\pi\phi) 
    \label{eq:radial_velocity}
\end{align}
where $K_x$ and $K_y$ are the two components of the RV, and $\gamma$ is systemic velocity. Table \ref{tab:rv_disk} shows the median values of all parameters and errors (16th, 84th quartiles).

The  $\rm H{\alpha}$ 6562.8 \AA\ and $\rm H{\beta}$ 4861.35 \AA\ Doppler tomograms, along with trailed spectra and RV curves, are shown in Figure \ref{fig:doppler}. Both Doppler tomograms show no accretion disk structure or bright spot. A prominent emission is located approximately 220 degrees close to WD and could be associated with possible stream accretion (see Section \ref{sec:discussion} and Appendix \ref{appendix:inv_dop} for discussion).

\subsection{Estimation of Binary Parameters}
\label{sec:estimation_BP}

\subsubsection{Distance, Extinction, and $N_H$}
\label{sec:distance_av}

The Gaia source associated with SRGeJ0411 within a 3.3$\arcsec$ search radius (98\% localisation error) has an ID 490854563672917120 (Gaia EDR3) and celestial coordinates  RA=$04^{h} 11^{m} 30^{s}.1$ and DEC=$+68\degr 53^{'} 49^{''}.9$. The corresponding Galactic coordinates are $\ell, b = 139.3574328\degr, 12.6797562\degr$. The distance calculated using Gaia EDR3 parallax is $d = 324^{+26}_{-31}$~pc \citep{2021bailerjones}. The 3-dimensional Bayestar19 dustmap of \cite{bayestar19} derives an extinction value of $E(B-V) = 0.13\pm 0.03$. We use the extinction law of CCM89 and a value of $R_V = 3.1$ to calculate the flux correction from UV to IR wavelengths. 
We can now calculate the intervening hydrogen column, $N_H = (8.9 \pm 3.3) \times 10^{20} \textrm{cm}^{-2}$, using the relation from \cite{2009MNRAS.400.2050G}.

\subsubsection{ SED modelling  }
\label{sec:sed}

The observed SED of SRGeJ0411 allows us to constrain the temperature and radius of the WD and donor. We construct the SED of SRGeJ0411 using photometry from GALEX \citep{2005galex} and  Pan-STARRS \citep{2016panstarrs}. We obtain all photometry by querying the respective tables within a $2\arcsec$ radius of the \textit{Gaia} position of SRGeJ0411 by using the Vizier database. The unWISE catalog \citep{2019ApJS..240...30S} provides four detections within a $5\arcsec$ radius search with different values of W$_1$ and W$_2$ photometry. This could suggest that there is possible contamination from nearby sources or that SRGeJ0411 is variable in the mid-IR (see Appendix \ref{appendix:wise} for a detailed discussion). We omit all WISE data from the SED approximation and place limits on the donor temperature from optical data alone.
\\

 {\it -- White Dwarf Temperature and Radius} 
\\

\begin{figure}
    \centering\includegraphics[width=0.48\textwidth]{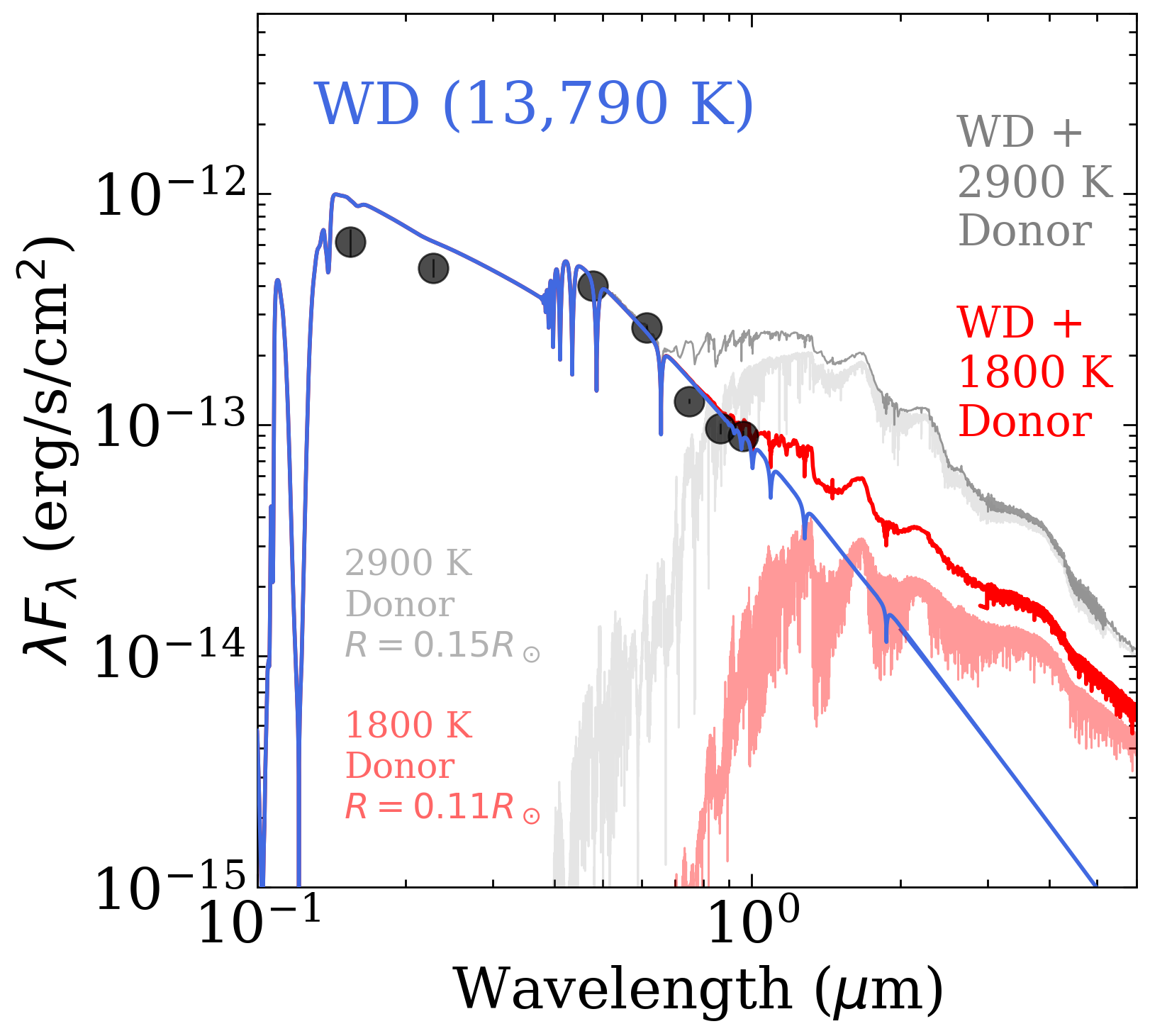}
    \caption{ The observed UV + optical SED of SRGeJ0411 is well fit by either a single 13,790 K temperature WD, or the sum of a 13,790 K temperature WD and \textit{at most} a 1,800 K temperature donor (3$\sigma$ upper limit). A pre-bounce (2,900 K) donor would significantly contribute in the red optical photometry, indicating that SRGeJ0411 is likely a period bouncer with a brown dwarf donor.}
    \label{fig:sed}
\end{figure}

We first approximate the SED as a hydrogen atmosphere (DA) WD constructed from Koester WD synthetic photometry \citep{2010koester}. We create a grid of models for the WD, with $T_\textrm{eff, WD}$ ranging from 10,000 K to 25,000 K. We fix the surface gravity of the WD at $\log g=$ 8.0. To constrain the radius of the WD, we multiply the model's surface fluxes by the factor of $\rm (R/d)^2$ to get the flux observed from the Earth. 

We perform two Bayesian analyses using the MCMC technique. The first analysis samples the posterior distribution of the WD radius and WD effective temperature, keeping distance and extinction fixed at their literature values (see Section \ref{sec:distance_av}). The second MCMC analysis additionally samples the posterior distribution of distance and extinction, $A_V$, taking the literature values as Gaussian priors ($d = 324\pm 30$ pc; $A_V = 3.1\times E(B-V) = 0.40\pm 0.1$). In both cases, we use an affine invariant sampler as implemented in \texttt{emcee} \citep{2013emcee}. We assume a Gaussian likelihood and set a uniform prior on $R_\textrm{WD}$ (0.005 -- 0.5 $R_\odot$) and a uniform prior $T_\textrm{eff, WD}$ (10,000 -- 25,000 K). We run the sampler for 10,000 steps, taking half as the burn-in period. 

The result of the first MCMC run, keeping distance and $A_V$ fixed, gives a rather high value of $\approx18,000$ K for the effective temperature of the WD (the $\chi^2$/dof value for this MCMC run is 18/5$\approx$3.6). Such a high temperature is unusual for WDs in period bouncers and any CV below the period gap \citep[e.g.][]{2011knigge, 2020belloni}. It is possible that this high temperature could be due to contribution from an accretion bright spot and/or accretion stream, whose modelling is beyond the scope of this paper.

The second MCMC run explores values of distance and extinction. The credible intervals (16th and 84th percentiles) of the marginalized posterior distribution of the distance agree with \cite{2021bailerjones}, but the extinction value is significantly different: $A_V \;(\textrm{MCMC}) = 0.03 \pm 0.03$ vs $A_V \;(\textrm{Bayestar19}) = 0.40 \pm0 .1$. The $\chi^2$/dof value for the MCMC routine exploring distance and extinction is 8.9/3$\approx$3.0. While acceptable as a first approximation, neither of the two models lead to an excellent fit, especially in UV photometry, likely due to the lack of modelling of an accretion disk, stream, or bright spot. From statistics alone, we cannot choose one model over the other. Since SRGeJ0411 is nearby enough and high enough above the Galactic plane to expect minimal extinction, we proceed with the value of $A_V = 0.03 \pm 0.03$ for the remainder of this paper. We use the mass-radius relation from \cite{2017massradius} to obtain an estimate of the WD mass. We emphasize that all parameters aside from the WD effective temperature have a weak dependence on the choice of $A_V$ value.
\\

{\it -- Donor Temperature and Radius: why the long period of SRGeJ0411 supports its period bouncer nature}
\\

MCMC runs with the donor temperature and radius being free parameters in the fit show unreasonably large donor radius ($\rm \gtrsim  0.6\ R_\odot$) and low donor temperature ($\rm \lesssim 1,000\ K$). This is likely caused by the fact that a WD model can sufficiently approximate the optical and UV photometry alone (blue line in Figure \ref{fig:sed}). We chose to then fix the donor radius, assuming a physically reasonable value of the donor radius to place an upper limit on the donor temperature.

We emphasize that SRGeJ0411 has an orbital period of 97.530 minutes, which is very far from the observed CV period minimum of 78 minutes. This means that there are two options for SRGeJ0411: either it is a pre-bounce system or a period bouncer. In Figure \ref{fig:sed} (grey colour), we show that a pre-bounce donor with an effective temperature of 2,900 K and radius of $0.15\ R_\odot$ is not a good fit for optical photometry. The Keck optical spectrum of SRGeJ0411 also does not present features for a pre-bounce CV (see Figure \ref{fig:LRIS}). Since SRGeJ0411 is unlikely to be a pre-bounce system, we see from the evolutionary tracks of \cite{2011knigge} that \textit{any} period bouncer should have a radius of $\lesssim 0.11R_\odot$ (see Section \ref{sec:comparison_tracks} and Figure \ref{fig:evolution_tracks}).  

We proceed with a third MCMC run, approximating the SED as a hydrogen atmosphere (DA) WD summed with a donor star, constructed from BT-DUSTY synthetic photometry \citep{2011btdusty}. We use the same grid of WD models and use a grid of donor models with $T_\textrm{eff, donor}$ ranging from 1,000 K to 3,000 K. We fix the surface gravity at $\log g=$ 5.0. Since the system is relatively nearby, we assume a solar metallicity for the donor, and we have no spectral information to infer a different value. We fix the donor radius at $0.11\ R_\odot$ and explore values of the donor effective temperature with a uniform prior between 1,000 and 3,000 K.  We obtain nearly identical values for all WD parameters and extinction. However, from this MCMC run, we obtain a 3$\sigma$ upper limit on the donor temperature of $T_\textrm{eff, donor} \lesssim 1,800$ K. Figure \ref{fig:sed} (red colour) presents the sum of a WD and 1,800 K donor model. It is clear that the reddest optical data (PanSTARRS $y$; $\approx 9600$ \AA) starts to severely disagree with any donor of a higher temperature. We note that if we choose to fix the radius at 0.15 $R_\odot$ (as in the pre-bounce donor), then we obtain a 3$\sigma$ upper limit of $\lesssim$1,600 K. This still suggests that SRGeJ0411 is a period bouncer with a cold brown dwarf donor.

\subsubsection{The Mass Ratio}
\label{sec:LC_and_q}

\begin{figure*}
    \centering
    \includegraphics[width=0.45\textwidth]{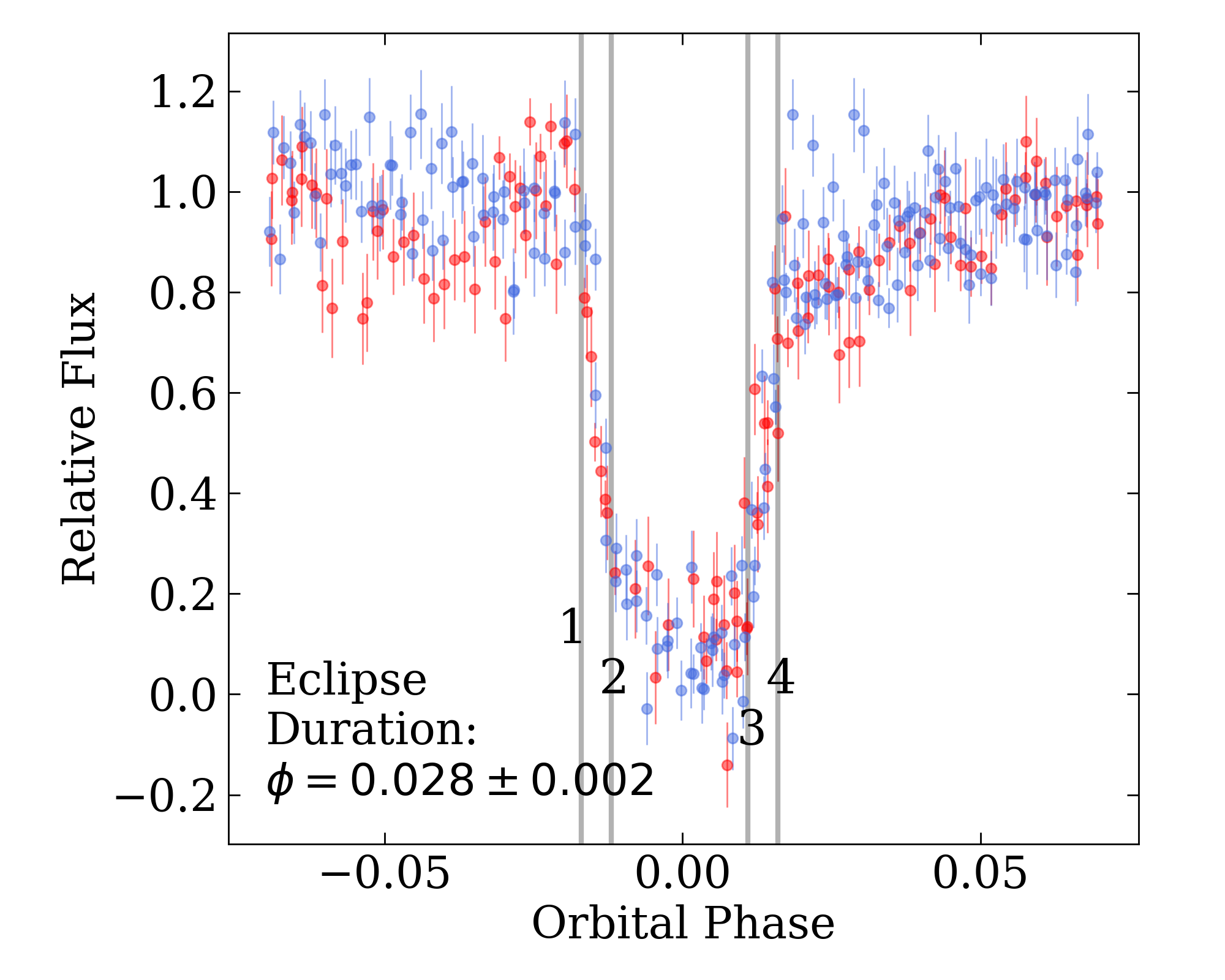}\includegraphics[width=0.4\textwidth]{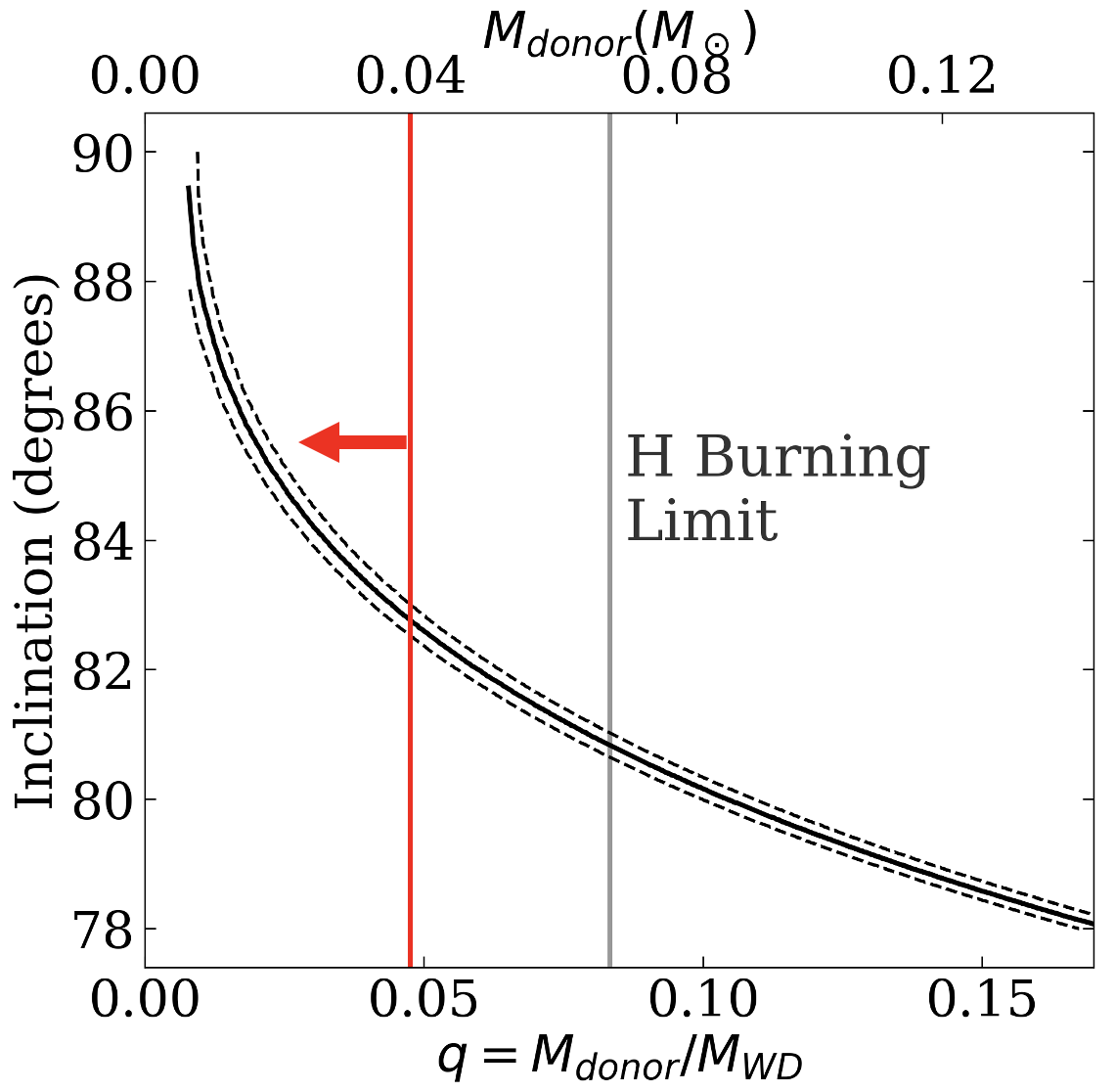}
    \caption{ {\it Left:} Phase-folded CHIMERA  $r$ ({\it red}) and $g$ ({\it blue}) high-speed optical light curve of SRGeJ0411 around the eclipse. {\it Vertical lines} correspond to the stages of the eclipse of the WD: first, second, third, and fourth contacts. {\it Right:} An eclipse time measurement constrains the inclination as a function of mass ratio, using the analytic equations of \protect\cite{1976ApJ...208..512C}. {\it The vertical red line} shows the upper limit on the donor mass computed from the period-density relationship for Roche-lobe filling stars. We note that this depends on the models of \protect\cite{2011knigge}, which place an upper limit on the radius of a period bouncer to be $\lesssim 0.11 R_\odot$. See Section \ref{sec:LC_and_q} for more details. } 
    \label{fig:lc_Channan}
\end{figure*}

To constrain the mass ratio of SRGeJ0411, we used the phase-folded CHIMERA light curve and the analytical relation from \cite{1976ApJ...208..512C}, which expresses the inclination as a function of mass ratio, given an eclipse time (the orbital phase length during which the donor eclipses the WD). This relation holds under the assumption that the donor is filling its Roche lobe. Therefore, we visually inspect the CHIMERA light curve to find the first, second, third, and fourth contacts between the donor and WD (see \cite{1984eclipse_model} for a definition of terms and example with a well-studied CV). We show the eclipse portion of the light curve, and points of contact in the left panel of Figure \ref{fig:lc_Channan}. We obtain an eclipse phase duration, or equivalently, the time between the mid-ingress and mid-egress of the primary, by fitting a trapezoidal eclipse profile to the ingress and egress of the eclipse. We estimate the phase duration between the primary mid-ingress and mid-egress to be $\phi=0.028\pm 0.002$, where the error corresponds to the 10 sec exposure time used in CHIMERA observations. This is equivalent to the difference between third and first or fourth and second contacts in Figure \ref{fig:lc_Channan}. 

From Chanan's equations and the eclipse time alone, we obtain a tight relation between the mass ratio and inclination (see the right panel of Figure \ref{fig:lc_Channan}). This constrains the range for the mass ratio: $0.02 \leq q \leq 0.15$, and the inclination angle: $78^\circ \leq i \leq 90^\circ$. The mass ratio suggests that the donor mass is in the $0.02 \leq M_{\rm donor} (M_{\odot}) \leq 0.13$ range. We compute the donor mass from the mass ratio by adopting the WD mass of $\rm 0.84\ M_\odot$ from SED modelling (see Section \ref{sec:sed}).  

We can more precisely constrain the mass of the donor star by considering the period-density relation for Roche-lobe-filling stars,
\begin{gather}
   \overline{\rho}_{\rm donor}= \frac{3M_{\rm donor}}{4\pi R_{\rm donor}^3} \approx 107\,{\rm g\,cm^{-3}}\left(P_{\rm orb}/{\rm hr}\right)^{-2},
    \label{eq:pho_donor}
\end{gather}
where $\overline{\rho}_{\rm donor}$ is the donor's mean density, and $M_{\rm donor}$ and $R_{\rm donor}$ are the mass and radius of the donor. The donor is assumed to have radius equal to its Roche lobe spherical equivalent radius $R_L$, that is computed based on \cite{1983ApJ...268..368E} approximation.

Using Eq. (\ref{eq:pho_donor}), the orbital period of SRGeJ0411, $P_{\rm orb}\approx 97.530$ minutes, and the donor radius $R_\textrm{donor} \lesssim 0.11\ R_\odot$ (see Section \ref{sec:sed}), we obtain the donor mass of $M_\textrm{donor} \lesssim 0.04 \ M_\odot$. 
The donor mass is typical for the known period bouncer CVs, and well below the maximum brown dwarf mass of $\approx0.07\ M_\odot$ \citep[e.g.][]{2019bd_mass}. We plot this value in the right panel of Figure \ref{fig:lc_Channan}. Returning to Chanan's equations, we constrain the inclination: $i(^\circ)\gtrsim 83$.

As an alternative to fixing the donor radius, we explore different possible values of inclination, which then set the value of mass ratio through the approximation of \cite{1976ApJ...208..512C}, and donor radius through the approximation of \cite{1983ApJ...268..368E}. We then fix the donor radius value (see MCMC run details in Section \ref{sec:sed}) and obtain upper limits on the donor effective temperature. For fixed inclination angles $i(^\circ)=(78, 83, 90)$, giving $q=(0.16, 0.04, 0.007)$ and $R_\textrm{donor}(R_\odot)=(0.15, 0.11, 0.07)$ respectively, we obtain a 3$\sigma$ upper limit of donor effective temperature $T_\textrm{eff, donor}(K)\lesssim$ (1800, 2100, 2250). Given our eclipse time, fixing any lower value of $\;i < 78^\circ$ would lead to donor temperatures colder than 1,800 K. Therefore, the donor temperature of SRGeJ0411 remains cold even for different inclination angles and agrees with the period bouncer scenario.

We note that we used Chanan's approximation to calculate the mass ratio of SRGeJ0411. We could see from the optical light curve that the pre-eclipse part of the light curve is distorted, which could be from a contribution of a bright spot (see Figure \ref{fig:CHIMERA}). For a more precise estimation of the mass ratio of components of SRGeJ0411, any emission from a bright spot, stream or accretion disk should be taken into account in the modelling of the phase folded light curve, but that work is beyond the scope of this paper. We emphasise that our assumptions place, at worst, upper limits on the mass ratio and donor mass, since an accretion disk and/or bright spot would only contribute to the overall flux level.

\subsubsection{ X-ray Spectrum, Luminosity and Mass Accretion Rate }
\label{sec:xray}

\begin{table}
\fontsize{8}{13}\selectfont
\centering
\caption{Results of approximation of  X-ray spectrum of SRGeJ0411 by different models. }
\label{tab:Xspectra}
\begin{tabular}{lcc}
\hline                     
\multicolumn{2}{l} { Model: ${\tt tbabs\times (powerlaw) }$ }           \\
\hline
{\it Parameters:}               &                                \\
$N_{\rm H}$($\times 10^{22}$cm$^{2}$)  & $0.04^{+0.04}_{-0.03}$   \\
$\Gamma$          & $1.19^{+0.27}_{-0.24}$     \\
$\rm C-stat$/(dof)$^{\rm a}$             & 24.6/37            \\
\hline     
\multicolumn{2}{l} { Model: ${\tt tbabs\times (mekal) }$ }           \\
\hline
{\it Parameters:}                  &                                \\
$N_{\rm H}$($\times 10^{22}$cm$^{2}$)  & $0.04^{+0.03}_{-0.02}$   \\
$kT_{\rm mekal}$(keV)            & $\gtrsim  8.5$     \\
$\rm C-stat$/(dof)$^{\rm a}$            & 24.7/32            \\
\hline
\hline
$F_{\rm 0.3-2.3\ keV}^{\rm } $  $\rm (\times 10^{-14}\ erg\ cm^{-2}\ s^{-1})$  & $19.0\pm 1.9$   \\
$L_{\rm 0.3-2.3\ keV}^{\rm } $ $\rm (\times 10^{30}\ erg\ s^{-1})$  & $2.4\pm0.2$   \\
\hline
\end{tabular}
\flushleft
\fontsize{7}{10}\selectfont
 Notes: (a) -- $C$-statistics and degree of freedom \citep{1979ApJ...228..939C}.
\end{table}

\begin{figure}
\centering
\includegraphics[width=0.48\textwidth,clip=true]{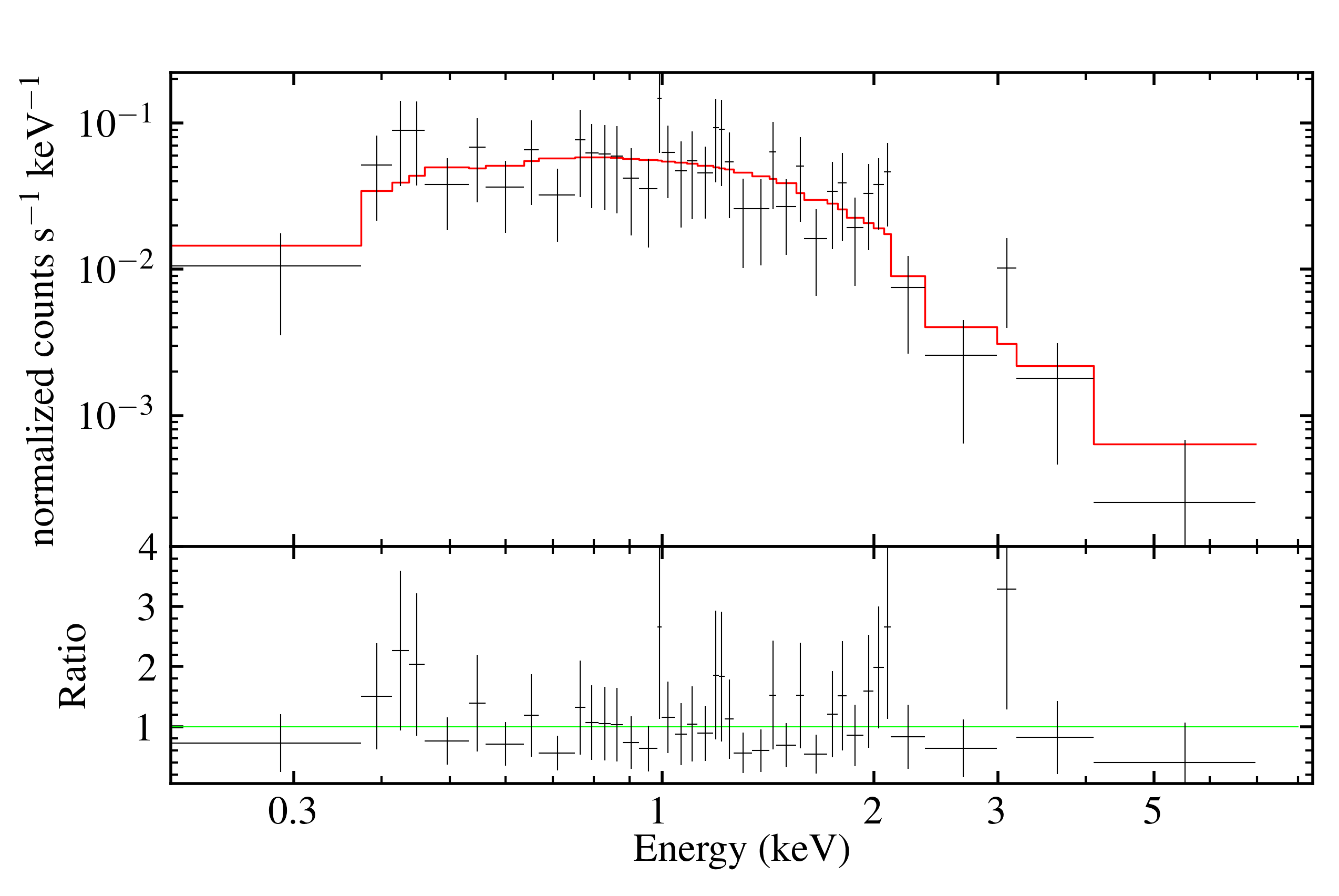}
\caption{The X-ray spectrum of SRGeJ0411 over four SRG/eROSITA all-sky surveys data ({\it top panel}). {\it The red line} shows the best-fit power-law model from Table \ref{tab:Xspectra}. {\it The bottom panel} shows the residuals (ratio of the data divided by the model) in each energy channel.}
\label{fig:Xray_spectrum}
\end{figure}

We approximated the combined X-ray spectrum of SRGeJ0411 within four all-sky survey data using two models: the power-law model ({\tt powerlaw} in XSPEC\footnote{XSPEC v.12 software \citep{1996ASPC..101...17A}}) and the optically thin thermal plasma model ({\tt mekal} in XSPEC) with solar abundance. We used the Tubingen-Boulder ISM absorption model ({\tt tbabs} in XSPEC, \citealt{2000ApJ...542..914W}) to account for interstellar absorption. The X-ray spectrum of SRGeJ0411 is shown in Figure \ref{fig:Xray_spectrum}, and the results of its approximation are shown in Table \ref{tab:Xspectra}. The hydrogen column density estimated from the approximation of the X-ray spectrum is low. It agrees with the hydrogen column density from the 3-dimensional Bayestar19 dustmap ($(8.9 \pm 3.3) \times 10^{20} \textrm{cm}^{-2}$, Section \ref{sec:distance_av}) and from SED modelling ($(6.6 \pm 6.6) \times 10^{19} \textrm{cm}^{-2}$, Section \ref{sec:sed}). Only a lower limit of the plasma temperature $\gtrsim$ 8.5 keV ($1\sigma$ confidence) can be estimated from the X-ray spectroscopy.

With its observed X-ray flux of $(19.0\pm 1.9) \times 10^{-14}\ \textrm{erg s}^{-1}\textrm{cm}^{-2}$ in the 0.3--2.3 keV energy band, SRGeJ0411 is not present in the Second ROSAT All Sky Survey Source Catalog, which had a flux limit of $\sim 2\times 10^{-13} \textrm{erg s}^{-1}\textrm{cm}^{-2}$ \citep[2RXS;][]{2016bollerA&A...588A.103B}. The absorption-corrected X-ray flux of SRGeJ0411 in the 0.3--2.3 keV energy band is  $(21\pm 1.3) \times 10^{-14} \textrm{ erg s}^{-1}\textrm{cm}^{-2}$, computed from the power-law model approximation. Given the well-constrained distance from \textit{Gaia}, the X-ray luminosity is $(2.6\pm 0.2) \times 10^{30} \textrm{ erg s}^{-1}$. To calculate the bolometric correction (BC) factor for X-ray luminosity in the 0.3--2.3 keV energy band, we used the canonical bremsstrahlung model. For the fixed temperature in the 8.5 -- 100 keV range, we compute the BC factor in the $\approx 3.3-14.8$ range. Taking this into account, we obtain an accretion rate of $\dot{M}=2L_{\rm X}R_{\rm WD}/M_{\rm WD}G \approx (1.7-7.8)\times10^{-12}$ $M_{\odot}$ yr$^{-1}$,  where $M_\textrm{WD}$ and $R_\textrm{WD}$ are the mass and radius of the WD from the results of SED modelling (see Section \ref{sec:sed} and Table \ref{tab:params}). The BC factor causes the accretion rate uncertainties.

\subsection{Comparison to CV Evolutionary Tracks}
\label{sec:comparison_tracks}

\begin{figure*}
\begin{center}
\includegraphics[width=0.85\textwidth]{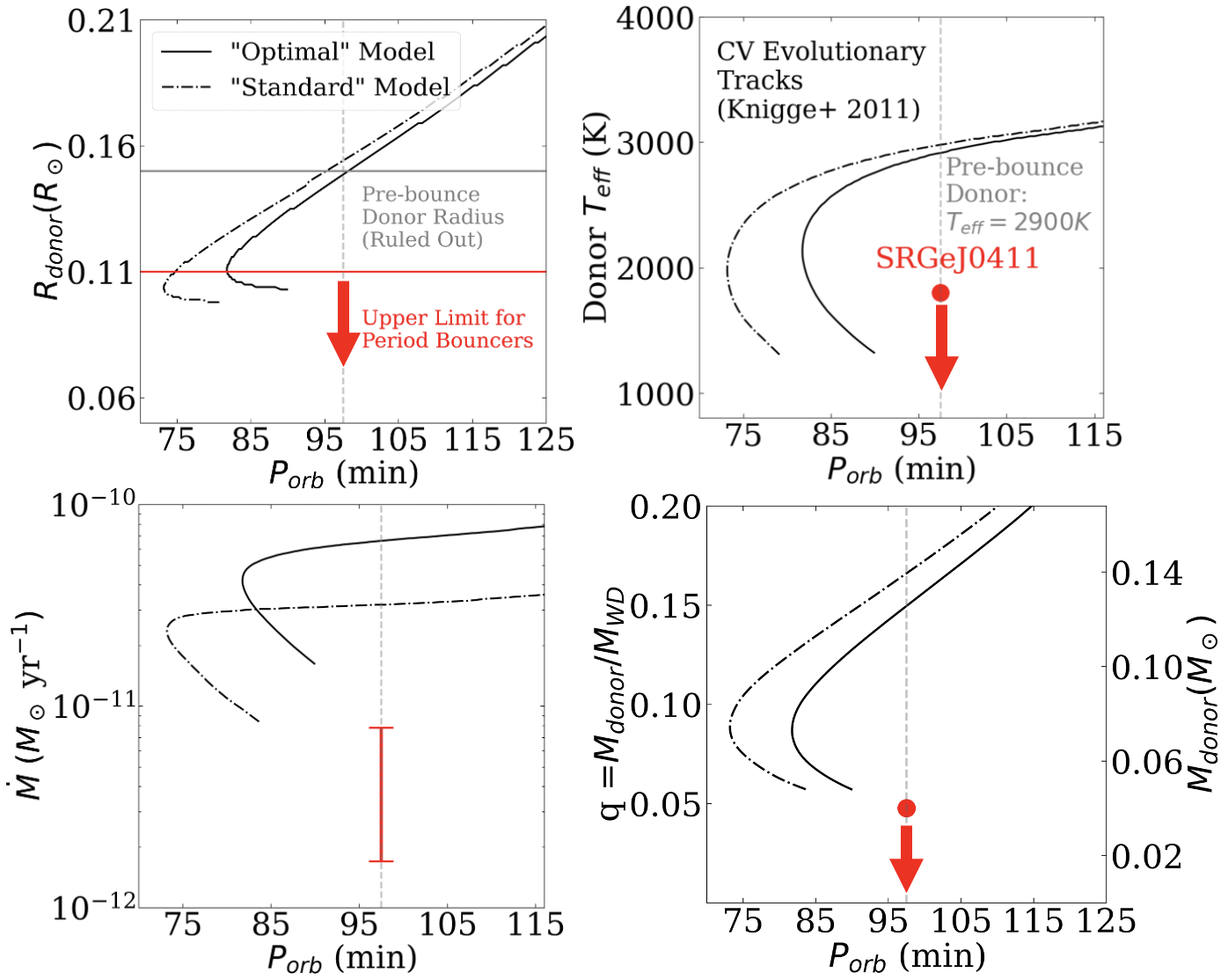}
\caption[]{ Comparison of observed properties of SRGeJ0411 with the CV evolutionary tracks from \protect\cite{2011knigge}. {\it Black lines} show the evolution of donor properties along the line for the "optimal" model ({\it solid line}) and "standard" model ({\it dash-dot line}) for CV evolution (see Tables 3--6, \protect\cite{2011knigge}). {\it Top panels}: donor radius and donor effective temperature as a function of orbital period. Because a pre-bounce donor can be ruled out from the SED and optical spectroscopy, we place upper limits on the donor radius, which allow us to place upper limits on the donor temperature and mass. {\it Bottom panels}: Mass-loss rate by the donor,  mass ratio of components, and donor mass as functions of orbital period. {\it Red markers} show the measured parameters (or parameter limits) of the SRGe0411 system from Table \ref{tab:params}.}
    \label{fig:evolution_tracks}
\end{center}
\end{figure*}

Table \ref{tab:params} summarises the binary parameters of SRGeJ0411 computed in previous sections (see Section \ref{sec:estimation_BP} for more details). We compared the observed properties of SRGeJ0411 with CV evolutionary tracks from \cite{2011knigge}. We note that \citet{2011knigge} computed CV evolutionary tracks only for one initial combination of binary parameters\footnote{We also note that \citet{2011knigge} assumed completely non-conservative evolution of the system.}. However, CV evolutionary tracks with different initial parameters are rather similar around the period minimum, so we used \citet{2011knigge} tracks to compare with our observational data.  Both the "optimal" (revised) and "standard" model tracks for CV evolution were used (see Tables 3--6, \cite{2011knigge}). 

Figure \ref{fig:evolution_tracks} shows the donor radius, effective temperature, accretion rate, mass ratio of the two components, and donor mass as a function of the orbital period. As explained in Section \ref{sec:sed}, there are two options for a CV with such a long period as that of SRGeJ0411: either it is a pre-bounce system ($T_\textrm{eff, donor} = 2,900$ K, $R_\textrm{donor} = 0.15\ R_\odot$) or it is a period bouncer well past the period minimum (period bouncers have an upper limit on their radii of $R_\textrm{donor} \lesssim 0.11\ R_\odot$). We show this reasoning in the upper left corner of Figure  \ref{fig:evolution_tracks}. 

After placing this upper limit on the donor radius, we used the SED analysis (Section \ref{sec:sed}) to place an upper limit on the temperature of the donor of $ T_\textrm{eff, donor}\lesssim 1,800$ K. The upper limit on the donor radius also allowed us to place upper limits on the donor mass using the Roche lobe filling period--density relation (Section \ref{sec:LC_and_q}). Both the donor temperature and donor mass show good agreement with the expected temperature and mass of period bouncers (upper and bottom right panels of Figure \ref{fig:evolution_tracks}). Finally, the X-ray luminosity provides an estimate of the accretion rate (lower left panel of Figure \ref{fig:evolution_tracks}; Section \ref{sec:xray}). The estimated low value of the accretion rate provides additional evidence (independent of any donor radius assumptions) that SRGeJ0411 is a period bouncer.

To sum up, all the computed parameters of SRGeJ0411 are consistent with the theoretical CV evolutionary tracks, suggesting that the SRGeJ0411 has passed through the period minimum during its evolution. We note that the WD mass of SRGeJ0411 is $M_\textrm{WD} = 0.84^{+0.07}_{-0.07}\ M_\odot$, which is consistent with the mean WD mass in CVs $0.81^{+0.16}_{-0.20}\ M_\odot$ \citep[e.g.][]{2022pala}, but higher than the mean mass of single WDs \citep[$\approx 0.6 M_\odot$;][]{2007kepler}. The estimated mass of the donor in SRGeJ0411 places the system close to or past the end of the CV evolutionary tracks of \citet{2011knigge}, which run for $\approx$3 Gyr (optimal) and $\approx$7 Gyr (standard). This means that these values may be considered as lower limits of the age of SRGeJ0411.

\section{Discussion}
\label{sec:discussion}

\begin{table}
\fontsize{8}{13}\selectfont
\centering
\caption{Binary Parameters of SRGeJ0411.}
\label{tab:params}
\begin{tabular}{lcc}
\hline                     
Parameter & Value & Origin\\
\hline
Distance, $d$ (pc) & $324^{+26}_{-31}$ & (1) \\
Orbital Period, $P_\textrm{orb}$ (min) & $97.529544\pm0.000173$ & (2)\\
\hline
Extinction, $A_V$ & $0.03^{+0.03}_{-0.03}$ & (3)\\
WD surface gravity, $\log g$  & 8.0 (fixed) & (3)\\
WD temperature, $T_\textrm{eff, WD}$ (K) & $13,790^{+530}_{-650}$ & (3)\\
WD radius, $R_\textrm{WD}$ ($0.01 R_\odot$) & $1.0^{+0.09}_{-0.09}$ & (3)\\
Donor surface gravity, $\log g$  & 5.0 (fixed) & (3)\\
Donor temperature, $T_\textrm{eff, donor}$ (K) & $\lesssim 1,800$ & (3)\\
Donor radius, $R_\textrm{donor}$ ($R_\odot$) & $\lesssim 0.11$ & (3)\\
\hline
WD mass, $M_\textrm{WD}$ ($M_\odot$) & $0.84^{+0.07}_{-0.07}$ & (4)\\
\hline
Accretion rate, $\dot{M}$ ($M_\odot \textrm{ yr}^{-1}$) & $ (1.7-7.8) \times 10^{-12}$& (5)\\
\hline
Mass ratio, $q$ & $\lesssim 0.05$ & (6) \\
Donor mass, $M_\textrm{donor}$ ($M_\odot$) & $\lesssim 0.04$  & (6) \\
Inclination, $i$ ($^\circ$) & $\gtrsim 83$ & (6) \\
\hline
\end{tabular}
\flushleft
\fontsize{7}{10}\selectfont
Notes: (1) -- Gaia parallax (Section \ref{sec:distance_av}); (2) -- Optical photometry (Section \ref{sec:period}); (3) -- SED (Section \ref{sec:sed}); (4) -- SED + WD mass-radius relation (Section \ref{sec:sed}); (5) -- X-ray (Section \ref{sec:xray}); (6) -- SED + Roche Lobe filling relation (Section \ref{sec:LC_and_q}).
\end{table}

\subsection{Single-peaked Emission Lines}

The single-peaked nature of the emission lines in SRGeJ0411 is one of the most puzzling aspects about the system (see Figure \ref{fig:LRIS}). During no phase do any of the spectra show line doubling as would be expected from an accretion disk \citep[e.g.][]{1980cv_disk_lines, hellierbook}. The lines are maximally redshifted in the middle of the eclipse ($\Phi$ = 0) and maximally blueshifted when the WD and donor are aligned with the WD in front of the donor ($\Phi$ = 0.5) (see Figure \ref{fig:doppler}). The single-peaked lines are also very broad --- the standard deviation of the Gaussian fit to the phase-averaged H$\alpha$  line profile is $\sigma\approx$ 330 km/s. Because this and other prominent emission lines span such a large range of velocities, they are unlikely to be due to irradiation of the donor. This also discards the possibility that it could be a bright spot on the accretion disk. Even if the bright spot were misaligned from the WD--donor line of sight and more consistent with the observed RV shifts, a small bright spot would not be sufficient to create such broad lines. These features, along with the rather large RV amplitude of the Balmer lines ($\approx 260$ km/s for H$\alpha$) are reminiscent of the broad emission lines seen in polars, including eclipsing systems \citep[e.g.][]{1990cropper}. 

\subsection{Unlikely SW Sex Classification}

In addition to polars, this particular combination of an eclipsing system, along with single peaked emission lines that are maximally redshifted at eclipse, is reminiscent of SW Sex stars \citep{1991thorstensen}. However, there are differences between SRGeJ0411 and SW Sex:  He II 4686 is absent in SRGeJ0411, unlike in SW Sex stars, and  SRGeJ0411 has a short duration, deep eclipses, unlike the broad triangular eclipses seen in SW Sex stars \citep{1991thorstensen}. Moreover, the presence of WD Balmer absorption lines in SRGeJ0411 suggests it is in a "low accretion state" (as defined for SW Sex stars and other CVs), which in SW Sex stars would imply that the emission lines should reveal an accretion disk in the Doppler tomogram \citep{2015sw_sex}. Because the emission lines do not trace a disk, it is unlikely that SRGeJ0411 is a CV under the SW Sex classification.

\subsection{Is SRGeJ0411 a Magnetic Period Bouncer?}

We speculate that SRGeJ0411 could be a magnetic CV. The Doppler tomogram shows no disk structure, and the large RVs suggest the emission is produced relatively close to the WD (see Figure \ref{fig:doppler} and Appendix \ref{appendix:inv_dop}). The emission could therefore be produced in an accretion stream funnelling material down to the surface of a magnetic WD. 

While strong He II 4686 is often an indicator of magnetism in CVs \citep[e.g.][]{1992silber}, weak or absent He II 4686 is typical in low-state polars and ``low-luminosity intermediate polars" (LLIPs). Low-state polars often lack He II 4686 altogether, but show strong cyclotron features in optical spectra \citep[e.g.][]{2007vogel}. Furthermore, LLIPs show EW ratios of He II 4686/H$\beta$ of 0.1 or even consistent with 0 during quiescence, but do not show cyclotron features, similar to SRGeJ0411 \citep[e.g. V445 And, V1460 Her, V1025 Cen;][]{2005V455And, 2017V1460Her, 2002V1025Cen}. About a dozen LLIPs have been identified in the literature\footnote{\url{https://asd.gsfc.nasa.gov/Koji.Mukai/iphome/catalog/llip.html}} \citep{2014llip}.

We do not see any cyclotron humps in the optical spectrum, as would be expected in a low-state polar. We can infer this is a low-state system since we see the WD absorption lines \citep[in high state systems the accretion continuum dominates and WD absorption lines are not seen, e.g.][]{1990cropper}. Furthermore, we do not see any Zeeman splitting of lines in the optical spectrum (see Figure \ref{fig:LRIS}), and can therefore place an upper limit for the possible magnetic field strength of SRGeJ0411. The resolution of DBSP is $\approx1.5$ \AA, but in the H$\alpha$ line and other strong emission lines, we cannot detect Zeeman splitting for separations lower than $\lesssim 20$ \AA\ due to the breadth of the emission lines. This corresponds to an upper limit for the magnetic field of $B\lesssim 1$ MG \citep{2003magnetic_sdss_wd}. However, we emphasize that this is only an approximate limit based on limited data, so further observations are needed to constrain the magnetic field strength. The wavelengths and strength of cyclotron features also depend on the viewing angle of the magnetic field \citep{2015ferrario}, which is unknown for SRGeJ0411. We are additionally limited by low signal-to-noise ratio in the continuum, which could prevent us from identifying weaker metal lines that could show Zeeman splitting. To sum up, our current understanding leads us to tentatively suggest that SRGeJ0411 is a magnetic period bouncer, but additional data is needed to confirm this.

\subsection{Clues from X-ray and Infrared Data}

SRGeJ0411 seems to have a fairly high X-ray luminosity compared to other intrinsically faint CVs, which could imply the WD is magnetic \citep[see e.g.,][]{2010MNRAS.408.2298B, 2013MNRAS.430.1994R}. X-ray spectroscopy also indirectly implies the magnetic nature of SRGeJ0411, where the photon index is steep $\Gamma\sim 1$ (see for discussion \cite{2021AstL...47..587G}). However, we require polarimetric studies or infrared spectroscopy to search for cyclotron features to reach any meaningful conclusion.

Infrared spectroscopy is needed to search for absorption and/or emission lines from the donor. In Appendix \ref{appendix:inv_dop}, we show the H$\alpha$ line from the Keck I/LRIS spectrum. A narrow ($\approx50$ km/s) component is seen in addition to the persistent broad ($\approx330$ km/s) component. Further evidence of irradiation of the brown dwarf by the X-ray and UV emission in the system could make SRGeJ0411 an interesting laboratory for the heating of brown dwarf atmospheres. Due to the low surface gravity of brown dwarfs, a strong wind could be induced, potentially leading to atmosphere evaporation \citep[e.g.][]{1977basko}. Infrared spectroscopy is the best tool to probe the nature of the brown dwarf and any heating/winds, although infrared photometry could also serve as a first step to search for differences in the day/night side temperatures. 

In Appendix \ref{appendix:wise}, we explore the WISE detections near SRGeJ0411 and whether they could indeed be associated with it or with blending from a nearby source. Taking WISE data into account, we carry forth with the suggestion that SRGeJ0411 is magnetic, and perform SED modelling assuming a WD, donor, and cyclotron emission. Further infrared photometry and/or spectroscopy as needed to confirm the nature of infrared emission from SRGeJ0411.

\section{Summary and Conclusion}
\label{sec:summary}

We have reported the discovery of a new eclipsing CV, SRGeJ0411, from a joint SRG/eROSITA and ZTF program, which shows evidence of being a period bouncer. Our results are summarized as follows:

\begin{itemize}

\item[--]  SRGeJ0411 has been discovered in the SRG/eROSITA all-sky survey, and it was identified as CV candidate through its high X-ray to optical flux ratio ($F_X/F_\textrm{opt} \approx 0.60$), Gaia parallax and optical colours. The observed X-ray luminosity of SRGeJ0411  is $L_X \approx 2.4\times 10^{30}\ \rm erg/s$ in the 0.3--2.3 keV energy band. SRGeJ0411 is a CV with a low  accretion rate $\dot{M} \approx (1.7-7.8)\times 10^{-12}\ M_\odot \textrm{ yr}^{-1}$.

\item[--] The Keck/LRIS optical spectrum shows prominent hydrogen and helium emission lines, typical for CVs (see Figure \ref{fig:LRIS}). Interestingly, we see only single-peaked emission lines and do not detect double-peaked lines as would be expected from the accretion disk of an eclipsing non-magnetic CV. No prominent emission or absorption lines from a donor star or disk, typically seen in CVs such as the Ca II triplet (8498, 8542, 8662  \AA), the Na I doublet (8183, 8195  \AA), or any TiO headbands seen in pre-bounce CV donors, were detected.

\item[--] The ZTF optical light curve shows deep eclipses ($\approx2.5^m$), and no significant outburst along a five-year-long baseline. The CHIMERA high-speed photometry light curves show a $\approx$20\% modulation in flux through excess brightness before the eclipse, possibly caused by a bright spot and/or accretion stream. The RTT-150 photometry reveals 
variability  in the light curve out of eclipse between observations, which could be caused by variations of accretion rate in the system. We estimate the orbital period of  SRGeJ0411 to be $P_\textrm{orb} = 97.529544\pm0.000173$ minutes from the combination of RTT-150 and P200/CHIMERA high-speed photometry. No significant X-ray variability was detected within four SRG/eROSITA all-sky surveys (see Figures \ref{fig:ztf_lc} -- \ref{fig:Xray_LC}).  

\item[--]  Table \ref{tab:params} presents all computed binary parameters of SRGeJ0411. The SED modelling shows that the donor has an effective  temperature $\rm \lesssim 1,800\ K$ consistent with the expected temperatures for donors of period bouncer CV (see Figure \ref{fig:sed}).  The comparison of the computed donor properties with the theoretical evolutionary tracks of CVs suggests that SRGeJ0411 passed the period minimum during its evolution (see Figure \ref{fig:evolution_tracks}).

\item[--]  The Doppler tomogram shows no disk structure, and the emission is produced relatively close to the WD (see Figures \ref{fig:doppler} and \ref{fig:inv-doptom}).  X-ray spectroscopy gives a relatively steep photon index $\Gamma\sim 1$, previously seen in magnetic CVs. This hints at the possible magnetic nature of SRGeJ0411, but further observations, including polarimetric studies, are required to get a meaningful conclusion.
\end{itemize}

\begin{figure}
    \centering
    \includegraphics[width=0.45\textwidth]{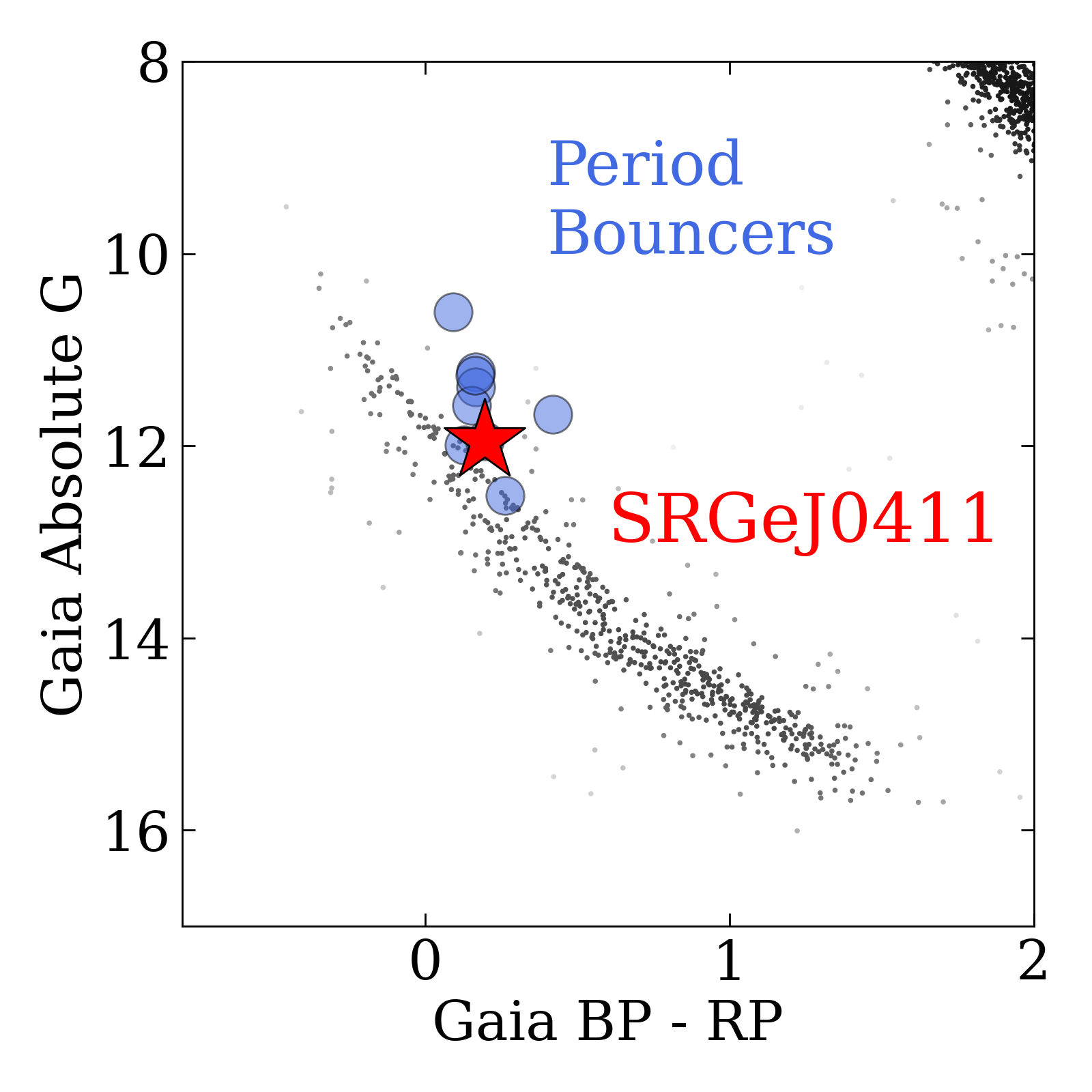}
    \caption{Position of SRGeJ0411 in the 100 pc Gaia Hertzsprung-Russell (HR) diagram \citep{2023A&A...674A...1G} alongside previously some known period bouncer systems \citep[see ][Table 5 and references therein]{2021ApJ...918...58A}  with a significant Gaia parallax ($\texttt{parallax\_over\_error} > 3$). Like other period bouncers, SRGeJ0411 lies slightly above the WD track. Searching for strong X-ray emitters in this area of the Gaia HR diagram could lead to the discovery of more period bouncers.}
    \label{fig:gaia_cmd}
\end{figure}
	
The multi-wavelength approach allows for detecting and investigating more CVs missed in one-band surveys alone. We plot the location of SRGeJ0411 in a 100 pc Gaia Hertzsprung-Russell diagram along with other known period bouncer systems in Figure~\ref{fig:gaia_cmd}, and show they are all located close to the WD track. This makes them faint and difficult to separate from isolated WDs with optical data alone. Recently, our joint program of searching CVs based on SRG/eROSITA and ZTF data led to the discovery of a similar system that can hide in the WD track: a 55-minute period eclipsing AM CVn, SRGeJ045359.9+622444 \citepalias{2023ApJ...954...63R}. In this work, our multi-wavelength analysis has led to the discovery of an eclipsing period bouncer system SRGeJ0411 due to it being an X-ray source in the SRG/eROSITA sky. This shows that combining multi-wavelength all-sky surveys allows for studying the single systems and Galactic CV populations.

\section*{Acknowledgements}

This work is based on observations with eROSITA telescope onboard SRG observatory. The SRG observatory was built by Roskosmos in the interests of the Russian Academy of Sciences represented by its Space Research Institute (IKI) in the framework of the Russian Federal Space Program, with the participation of the Deutsches Zentrum für Luft- und Raumfahrt (DLR). The SRG/eROSITA X-ray telescope was built by a consortium of German Institutes led by MPE, and supported by DLR. The SRG spacecraft was designed, built, launched and is operated by the Lavochkin Association and its subcontractors. The science data are downlinked via the Deep Space Network Antennae in Bear Lakes, Ussurijsk, and Baykonur, funded by Roskosmos. The eROSITA data used in this work were processed using the eSASS software system developed by the German eROSITA consortium and proprietary data reduction and analysis software developed by the Russian eROSITA Consortium.

Based on observations obtained with the Samuel Oschin Telescope 48-inch and the 60-inch Telescope at the Palomar Observatory as part of the Zwicky Transient Facility project. ZTF is supported by the National Science Foundation under Grants No. AST-1440341 and AST-2034437 and a collaboration including current partners Caltech, IPAC, the Weizmann Institute of Science, the Oskar Klein Center at Stockholm University, the University of Maryland, Deutsches Elektronen-Synchrotron and Humboldt University, the TANGO Consortium of Taiwan, the University of Wisconsin at Milwaukee, Trinity College Dublin, Lawrence Livermore National Laboratories, IN2P3, University of Warwick, Ruhr University Bochum, Northwestern University and former partners the University of Washington, Los Alamos National Laboratories, and Lawrence Berkeley National Laboratories. Operations are conducted by COO, IPAC, and UW. The ZTF forced-photometry service was funded under the Heising-Simons Foundation grant No.12540303 (PI: Graham).

We are grateful to the staffs of the Palomar and Keck Observatories for their work in helping us carry out our observations.  We thank T\"UB\.{I}TAK, the Space Research Institute of the Russian Academy of Sciences, the Kazan Federal University, and the Academy of Sciences of Tatarstan for their partial support in using RTT-150 (Russian - Turkish 1.5-m telescope in Antalya).

This work has made use of data from the European Space Agency (ESA) mission {\it Gaia} (\url{https://www.cosmos.esa.int/gaia}), processed by the {\it Gaia} Data Processing and Analysis Consortium (DPAC, \url{https://www.cosmos.esa.int/web/gaia/dpac/consortium}). Funding for the DPAC has been provided by national institutions, in particular the institutions participating in the {\it Gaia} Multilateral Agreement.

This research made use of matplotlib, a Python library for publication quality graphics \citep{Hunter2007}; NumPy \citep{Harris2020}; Astroquery \citep{2019AJ....157...98G}; Astropy, a community-developed core Python package for Astronomy \citep{2013A&A...558A..33A, 2018AJ....156..123A}; and the VizieR catalogue access tool, CDS, Strasbourg, France. The authors wish to thank E. Kotze for making his Doppler tomography code, \texttt{doptomog}, public \citep{2015kotze}.

IG acknowledges support from Kazan Federal University. ACR acknowledges support from the National Science Foundation via an NSF Graduate Research Fellowship. The work of IB, MG, IKh, AS, PM, RG supported by the RSF grant N 23-12-00292. LY gratefully acknowledges discussions with G. Tovmassian. BG acknowledges funding from the European Research Council (ERC) under the European Union’s Horizon 2020 research and innovation programme (Grant agreement No. 101020057). The authors thank the anonymous referee for their helpful comments and suggestions that improved the quality of the manuscript.

\section*{Data Availability}
X-ray data analysed in this article were used with the permission of the Russian SRG/eROSITA consortium. The data will become publicly available as a part of the corresponding SRG/eROSITA data release along with the appropriate calibration information. The Keck/LRIS, DBSP, CHIMERA and RTT-150 data underlying this article will be shared on reasonable request to the corresponding author. All other data are publicly available and can be accessed at the corresponding public archive servers.



\bibliographystyle{mnras}
\bibliography{EclipsingPBmnras} 




\appendix

\section{RV Curves and Inverse Doppler Tomogram: Additional Evidence of Magnetic Nature?}
\label{appendix:inv_dop}

\begin{figure*}
    \centering
    \includegraphics[width=0.3\textwidth]{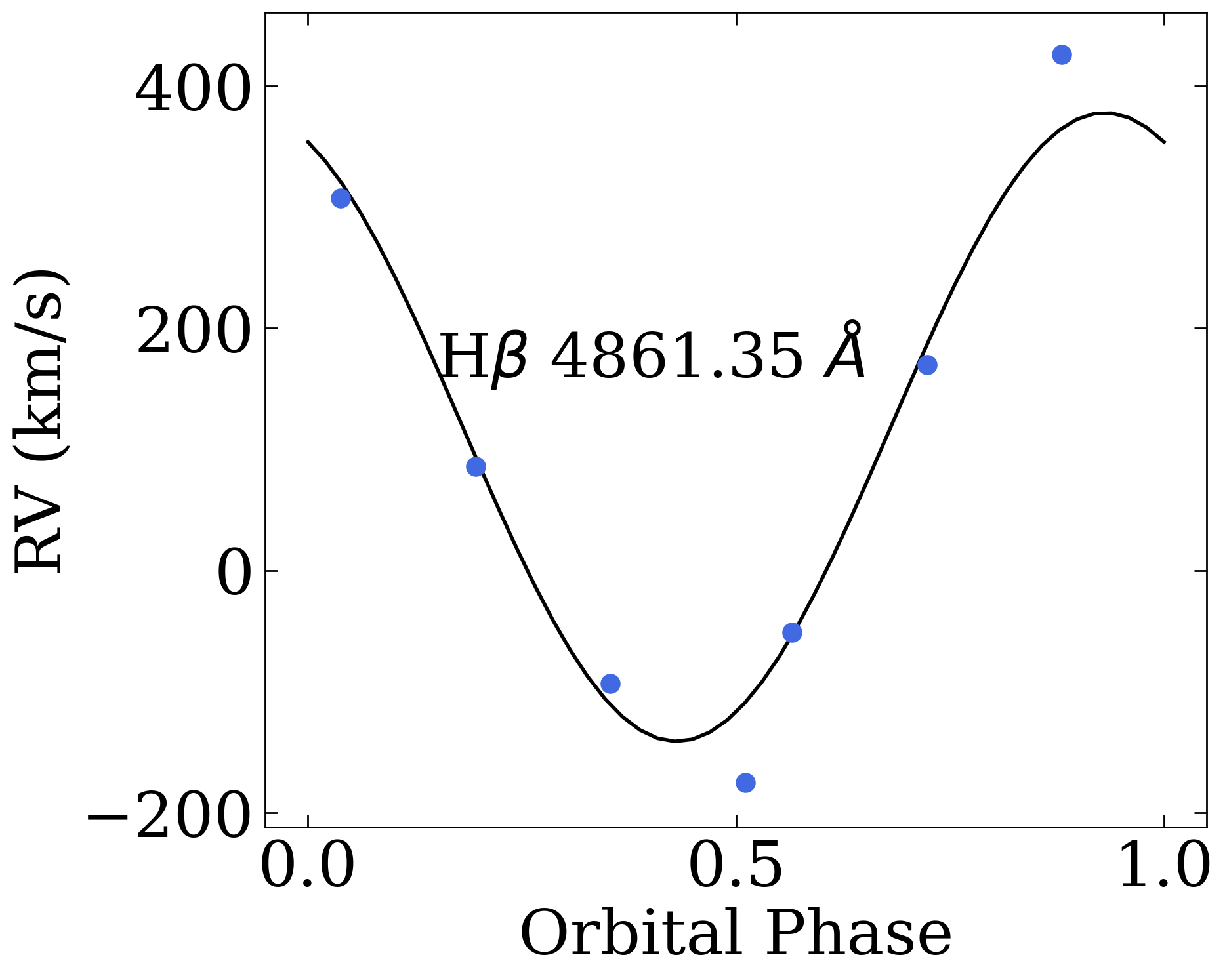}\includegraphics[width=0.3\textwidth]{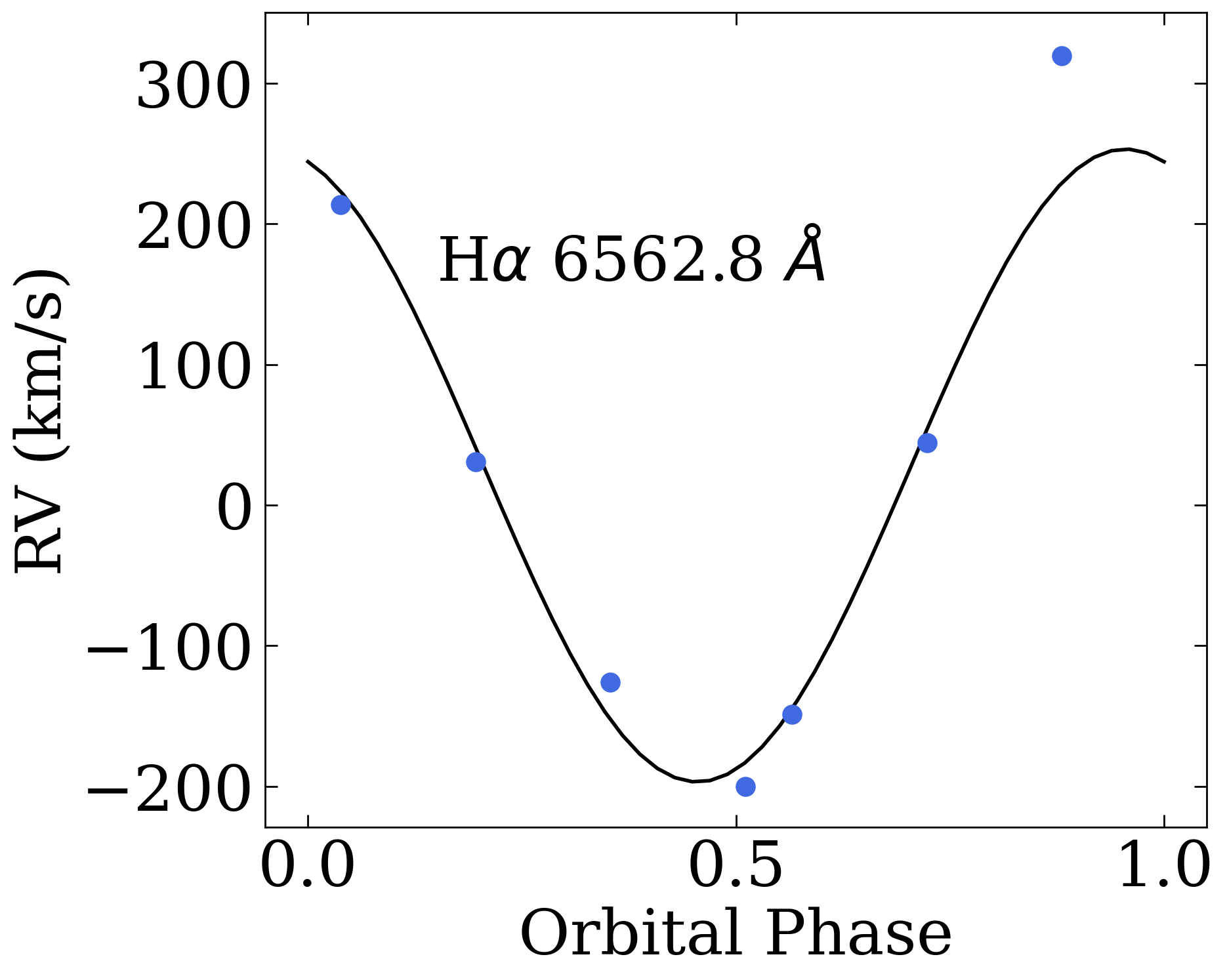}\includegraphics[width=0.3\textwidth]{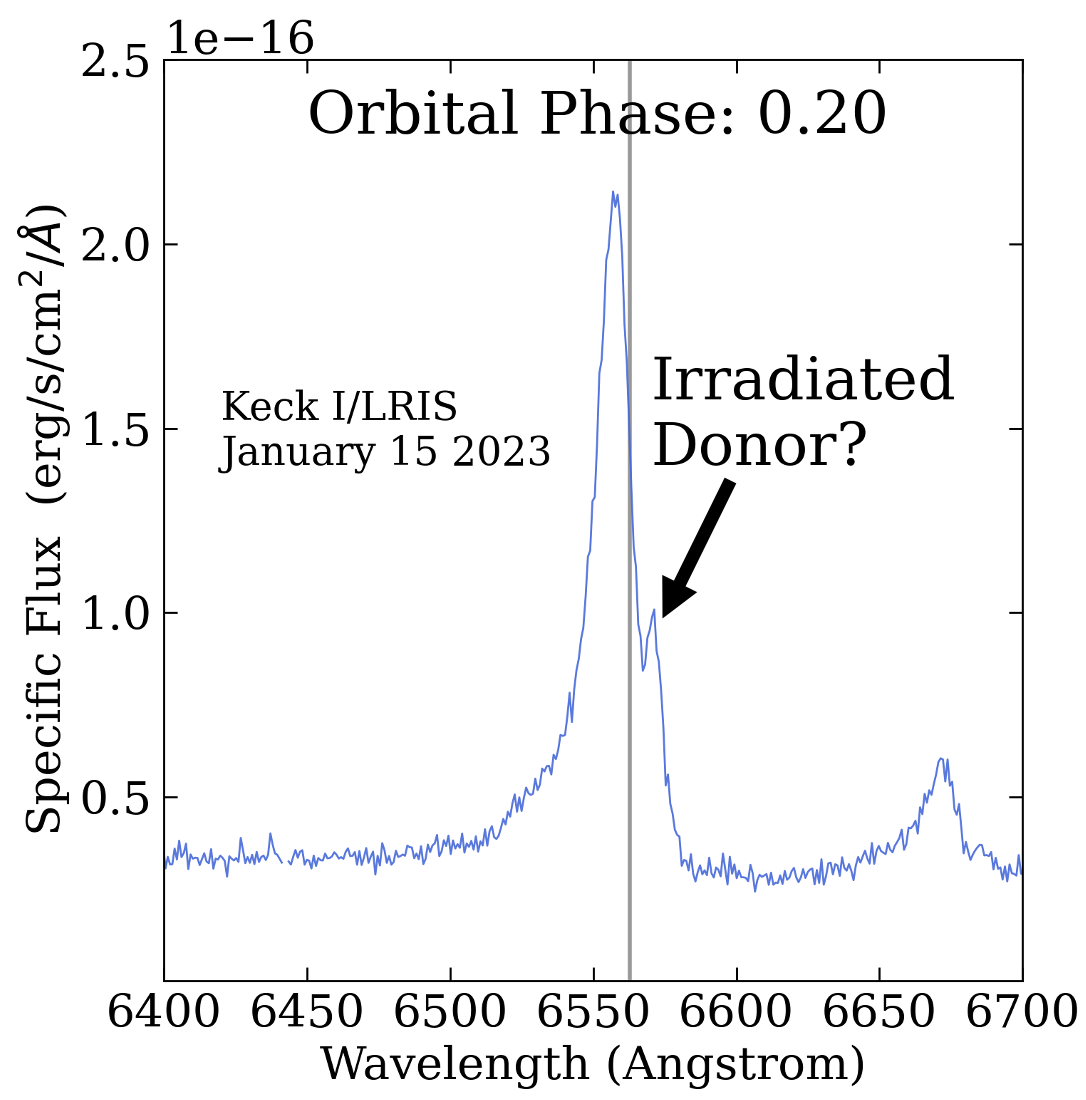}
    \caption{\textit{Left and center panels}: RV curves from P200/DBSP for H$\beta$ and H$\alpha$ lines, respectively. The RV amplitude is high ($\approx250$ km/s, see Table \ref{tab:rv_disk}), the lines are maximally redshifted when the donor eclipses the WD ($\Phi=0$), and maximally blueshifted half an orbital phase apart ($\Phi=0.5$). \textit{Right panel:} The only evidence of line doubling in SRGeJ0411 is seen in the H$\alpha$ line in a Keck I/LRIS spectrum taken at orbital phase $\Phi=0.2$. The narrow component is redshifted, while the main, broad component is blueshifted.}
    \label{fig:rv_curves}
\end{figure*}

\begin{figure}
    \centering
    \includegraphics[width=0.4\textwidth]{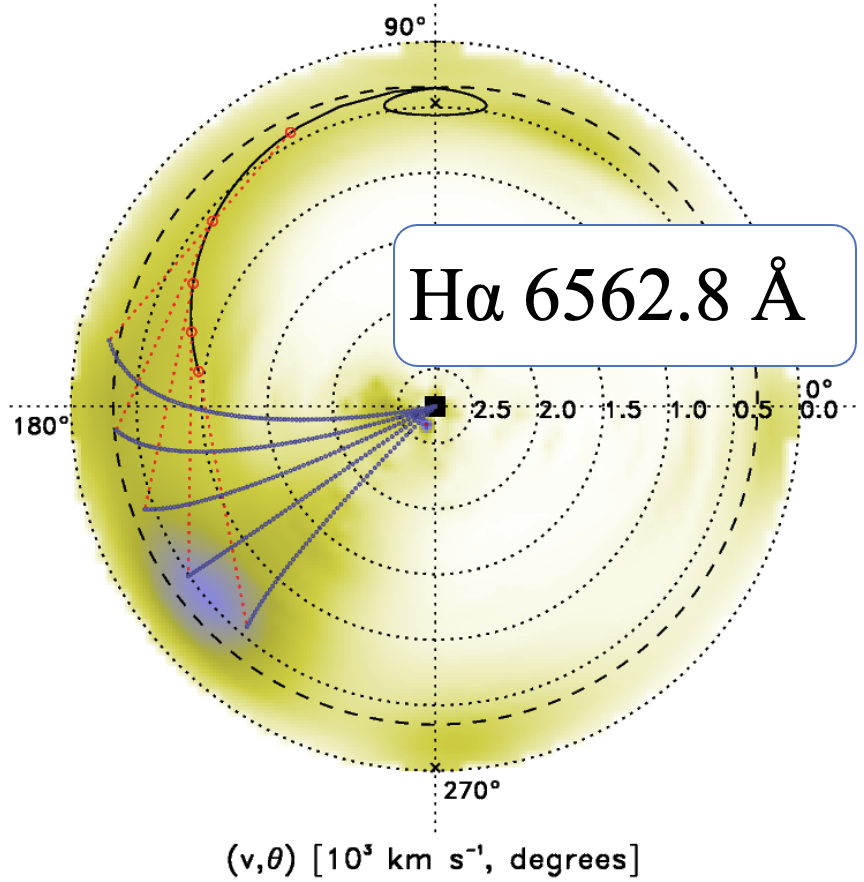}
    \caption{An inverse Doppler tomogram shows the absence of an accretion disk and concentrated emission within a small azimuthal range. Overplotted lines corresponding to the accretion stream and magnetic connection appear to be consistent with the emission but have model assumptions that require further evidence to confirm. The {\it thin dotted lines} with numbers indicate curves of equal velocity in km/s, and the {\it thick dotted line} indicates the Roche lobe of the WD. {\it The solid black curved line} indicates the path of an accretion stream for the mass ratio of SRGeJ0411, with material magnetically threading ({\it red lines}) onto {\it the solid grey line}, which represents a ballistic path onto the WD surface.}
    \label{fig:inv-doptom}
\end{figure}

In the left two panels of Figure \ref{fig:rv_curves}, we present the RV curves for the H$\beta$ and H$\alpha$ lines. Their high amplitude ($\approx 250$ km/s) is clear, along with their maximal redshift when the donor eclipses the WD and maximal blueshift when the WD is in front of the donor. This behaviour could be explained if the lines originate in a stream of material magnetically channelled towards the WD, as seen in polars. In the right panel of Figure \ref{fig:rv_curves}, we show the only instance of line doubling in any of the spectra of SRGeJ0411. Line doubling is not present in any of the P200/DBSP spectra but is present in this single spectrum taken with Keck I/LRIS, presumably due to the combination of increased signal-to-noise and slightly better resolution. A narrow ($\approx 50$ km/s) component is redshifted, while the broad ($\approx 330$ km/s) component is blueshifted, suggesting that the narrow component could originate on the face of the irradiated donor star. This is consistent with the spectrum having been taken at orbital phase $\Phi=0.2$, where the lines from the donor would be redshifted in orbit. 

We present an "inverse" Doppler tomogram in Figure \ref{fig:inv-doptom}. Whereas "regular" Doppler tomography in Figure \ref{fig:doppler} shows higher velocities at larger radii, inverse diagrams present the inverse in order to highlight higher velocities commonly seen in magnetic CVs \citep{2015kotze}. In Figure \ref{fig:inv-doptom}, we over plot the donor, ballistic stream trajectory, and accretion stream atop the inverse Doppler tomogram. For the purposes of this plot, we assume the system is a {\it polar} since the spectra across an entire orbit are dominated by broad, single-peaked emission lines with large amplitudes, characteristic of polars \citep{1990cropper}.

We use the upper limit of the mass ratio and the lower limit of the inclination from Table \ref{tab:params}. In order to create the accretion trajectories, we assume the azimuthal angle at which the accretion stream ends (45$^\circ$), the azimuthal angle at which the magnetic connection starts (5$^\circ$), and the azimuthal angle at which the magnetic connection ends (45$^\circ$). These angles correspond to the positions of the purple lines in the third quadrant in Figure \ref{fig:inv-doptom}. We refer the reader to \cite{2015kotze} for a more thorough description of these parameters. We have no way to determine the accretion geometry (the inverse Doppler tomograms we show are independent of magnetic field strength) and simply selected a combination of typical parameters (stated above) which happen to provide an adequate fit by-eye to the data in the Doppler tomogram.

Figure \ref{fig:inv-doptom}, like the normal Doppler tomograms in Figure \ref{fig:doppler}, shows the absence of an accretion disk spanning the entire azimuthal range. Emission is preferentially shown along the trajectory that an accretion stream would have, and the strongest emission is present at the point where the ballistic stream ends and the magnetic connection begins. While this picture is consistent with genuine polars in the literature, we remind the reader that further evidence is needed to determine the magnetic nature of SRGeJ0411.

\section{Possible hints from infrared WISE data for SRGeJ0411 being a  period bouncer}
\label{appendix:wise}

\begin{figure*}
    \centering
    \includegraphics[width=0.80\textwidth]{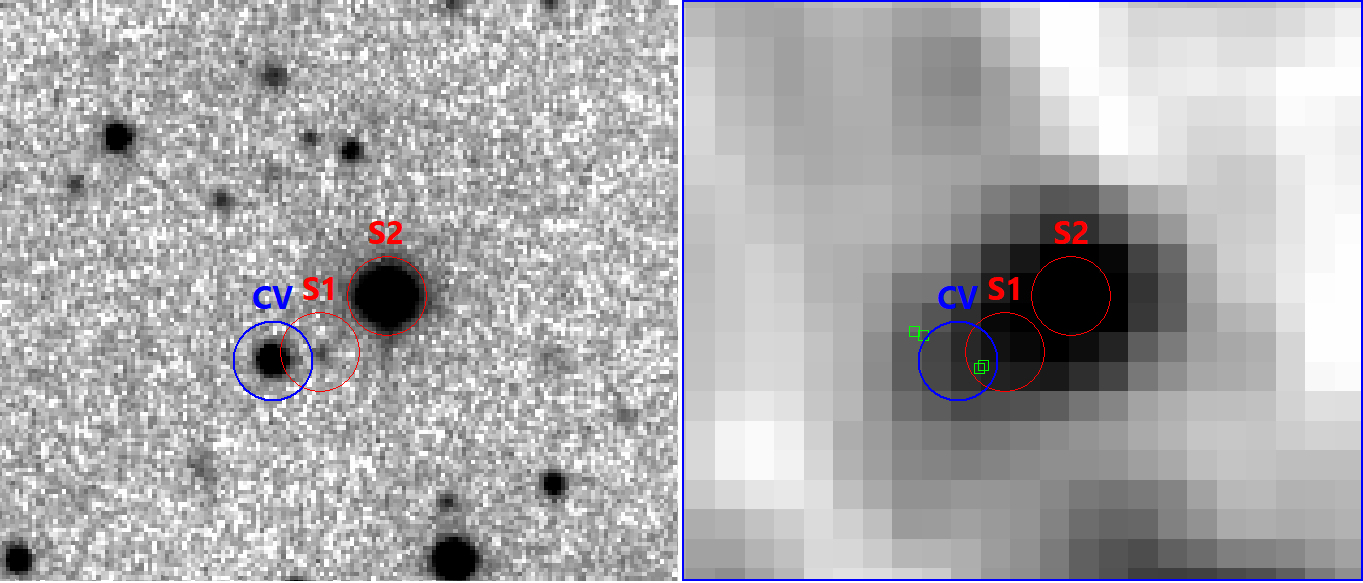}
    \caption{{\it Left:} Composite optical $griz$ image of SRGeJ0411 and nearby sources based on RTT-150 data. {\it Blue} circle shows SRGeJ0411 (CV), {\it red} circles show nearby sources (S1 and S2). Radii of circles are $3\arcsec$. {\it Right:} WISE image in W$_2$ filter. {\it Green squares} show positions of infrared sources from unWISE catalog near SRGeJ0411.}
    \label{fig:rtt_wise}
\end{figure*}

\begin{figure*}
    \centering
    \includegraphics[width=0.4\textwidth]{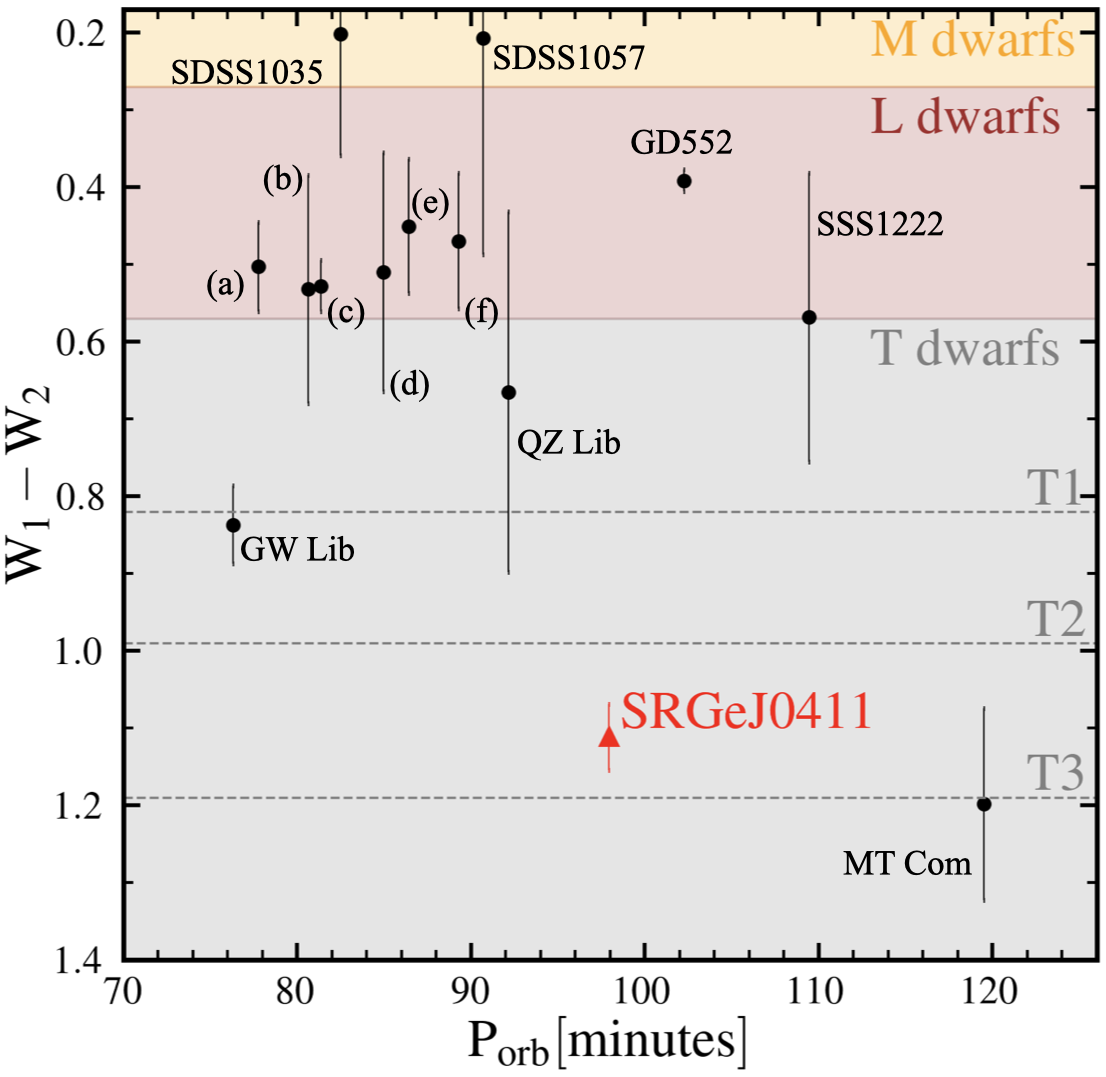}\includegraphics[width=0.565\textwidth]{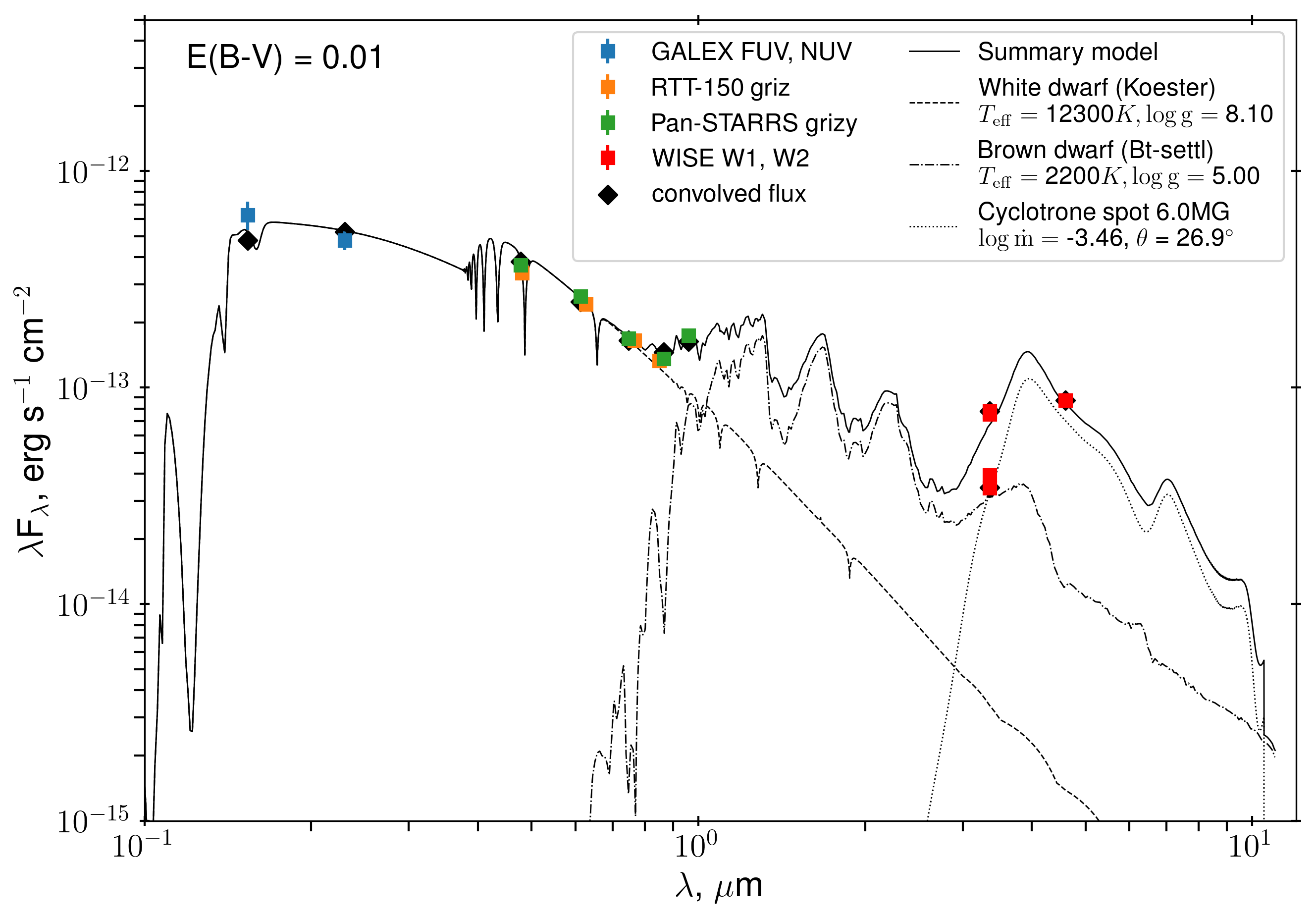}
    \caption{ {\it Left:} Infrared WISE $W_1-W_2$ colors versus the orbital periods of the period bouncer CVs. Magnitudes are in the Vega system. Labels show known period bouncers: (a) BW Scl; (b) V406 Vir; (c) V455 And; (d) EZ Lyn; (e) EG Cnc and (f) LP 731-60. {\it Red triangle} shows the SRGeJ0411 and {\it black dots} show some known period bouncer CVs \citep[see Table 5 and references therein,][]{2021ApJ...918...58A}. {\it Color regions} show the range of WISE colors computed for dwarfs templates with different spectral types \citep{2015A&A...574A..78S}, and {\it doted, horizontal lines} correspond to spectral types T1--T3. {\it Right}: SED of SRGeJ0411 with different photometry data. Lines correspond to different models: a WD atmosphere ({\it dashed});  a brown dwarf ({\it dash-dotted}); cyclotron spot ({\it dotted}).} 
    \label{fig:colors_sed}
\end{figure*}

Figure \ref{fig:rtt_wise} shows composite optical RTT-150 and an infrared WISE image of SRGeJ0411. The optical image shows  two resolved sources (S1 and S2) near SRGeJ0411, but in the WISE image, sources are blended  due to the large PSF size (FWHM $\approx6\arcsec$).  The unWISE catalog provides four individual sources within a $5\arcsec$ radius search near an SRGeJ0411 position with a high signal-to-noise ratio (above 5$\sigma$). These sources could be associated with nearby sources or be due to the possible infrared variability of SRGeJ0411. We inspect the WISE data, assuming that the four WISE sources are associated with SRGeJ0411, and show that they still could support the claim that SRGeJ0411 is a period bouncer.

Two sources from unWISE catalog ({\tt 0627p696o0047624} and {\tt 0632p681o0047052}) have almost similar (within 5\% variation) W$_1$ (3.4 $\mu m$) and W$_2$ (4.2 $\mu m$) magnitudes. The other two sources ({\tt 0627p696o0121297} and {\tt 0632p681o0119844}) show  only W$_1$ (3.4 $\mu m$) magnitudes and about 10\% variation between two measurements. We computed $\rm W_1-W_2$ based on the first two WISE sources and assumed that they were associated with SRGeJ0411. The {\it possible} observed WISE color of SRGeJ0411 is $\rm W_1-W_2 \approx1.1$. Left panel of Figure \ref{fig:colors_sed} shows $W_1-W_2$ colors versus the orbital periods of the known period bouncer CVs. The extinction coefficient ($A_{\lambda}$) is almost zero at long wavelengths; therefore, no correction was applied for infrared colors. The list of a dozen period bouncers with known orbital periods was adopted from \citet{2021ApJ...918...58A} (see Table 5 and references therein).  Colourized regions in Figure \ref{fig:colors_sed} (left panel) correspond to different colour ranges computed for M5 -- T8 dwarfs templates \citep[see Table 1,][]{2015A&A...574A..78S}. SRGeJ0411 occupies the same region as other known period bouncers in Figure \ref{fig:colors_sed}. The {\it possible} $\rm W_1-W_2$ color of SRGeJ0411 indicates that the donor of SRGeJ0411 is cold with an effective temperature of about $\sim$ 1,500 K (template colors of dwarfs of spectral type T2--T3 are $\rm W_1-W_2 =0.99-1.19$, \citet{2015A&A...574A..78S}), which agrees with the picture that the SRGeJ0411 could be a period bouncer CV.  

We discuss that the SRGeJ0411 could be a magnetic system in Section \ref{sec:discussion}. We performed an additional SED approximation\footnote{The GALEX (FUV, NUV), Pan-STARRS (grizy), RTT-150 (griz) fluxes were recalculated from AB magnitudes with correction for interstellar absorption (E(B-V) = 0.01). Fluxes from all four WISE sources were included in the SED approximation.} to model a WD, donor, and cyclotron emission source (see Figure \ref{fig:colors_sed}, right panel). We fixed the magnetic field and the angle between the magnetic field and the line of sight at 6 MG and $\theta=26.9\degr$, respectively, assuming that SRGeJ0411 is an intermediate polar. We model the fluxes using WD \citep{2010koester} and brown dwarf BT-Settl \citep{2012RSPTA.370.2765A} atmosphere models. The parameters for the WD and brown dwarf varied in the range $\rm T_\textrm{eff, WD}$ = 10,000 - 20,000 K, $\rm \log g$ = 8.0 (fixed) and $\rm T_\textrm{eff, donor}$ = 900 -- 2,500 K, $\rm \log g$ = 5.0 (fixed), respectively. As a result, we got an effective temperature of WD $\rm T_\textrm{eff, WD} \approx 12,300\ K$, and a brown dwarf $\rm T_\textrm{eff, donor} \approx 2,200\ K$. A WD has a radius and mass of $\rm R_\textrm{WD} \approx 0.011\ R_\odot$ and $\rm M_{WD} \approx 0.71\ M_\odot$ respectively, and a brown dwarf has a radius of $\rm R_{donor} \approx 0.13\ R_\odot$. Computed parameters are consistent with the results from SED modelling based on optical data alone (see Table \ref{tab:params}). If we assume that the cyclotron emitting area is circular, a radius is equal to $\rm R_{spot} \approx 2.5\times10^{-3}\ R_\odot$, and the ratio of the spot area to the surface area of the white dwarf is about $\rm S_{spot}/S_{wd} \approx 0.014$, which is consistent with the theory of magnetic accretion in the polar. The resulting SED approximation gives an accretion rate of about $\dot{M}$ $\rm \approx 6 \times 10^{-13}\ M_\odot \textrm{ yr}^{-1}$. 

We note that these results should be considered with caution. Without infrared spectroscopy, we cannot  conclude with certainty that the WISE colors are associated with SRGeJ0411, and if there is a cyclotron or even another source of emission. Future observations are required to investigate the infrared emission of SRGeJ0411. 


\bsp	
\label{lastpage}
\end{document}